\documentclass[fleqn,usenatbib]{mnras}

\usepackage{newtxtext,newtxmath}

\usepackage[T1]{fontenc}

\usepackage{graphicx}	
\usepackage{amsmath}	
\usepackage{color}
\usepackage{multirow}

\newcommand{\dd}[1]{{\mathrm{d}{#1}}}

\title[Galaxy merger rates with {\normalfont \textsc{Emerge}}]{{\normalfont \textsc{Emerge}}: Empirical predictions of galaxy merger rates since $z\sim6$}

\author[J. A. O'Leary et al.]{
Joseph A. O'Leary,$^{1}$\thanks{E-mail: joleary@usm.lmu.de}
Benjamin P. Moster,$^{1,2}$
Thorsten Naab,$^{2}$
Rachel S. Somerville$^{3}$
\\
$^{1}$Universit{\"a}ts-Sternwarte, Ludwig-Maximilians-Universit{\"a}t M{\"u}nchen, Scheinerstr. 1, 81679 M{\"u}nchen, Germany\\
$^{2}$Max-Planck Institut f\"ur Astrophysik, Karl-Schwarzschild Stra{\ss}e 1, 85748 Garching, Germany\\
$^{3}$Center for Computational Astrophysics, Flatiron Institute, 162 5th Avenue, New York, NY 10010, US
}

\date{Accepted XXX. Received YYY; in original form ZZZ}

\pubyear{2020}

\begin{document}
\label{firstpage}
\pagerange{\pageref{firstpage}--\pageref{lastpage}}
\maketitle

\begin{abstract}
	We explore the galaxy-galaxy merger rate with the empirical model for galaxy formation, \textsc{emerge}. On average, we find that between $2$ per cent and $20$ per cent of massive galaxies ($\log_{10}(m_{*}/M_{\odot}) \geq 10.3$) will experience a major merger per Gyr. Our model predicts galaxy merger rates that do not scale as a power-law with redshift when selected by descendant stellar mass, and exhibit a clear stellar mass and mass-ratio dependence. Specifically, major mergers are more frequent at high masses and at low redshift. We show mergers are significant for the stellar mass growth of galaxies $\log_{10}(m_{*}/M_{\odot}) \gtrsim 11.0$. For the most massive galaxies major mergers dominate the accreted mass fraction, contributing as much as $90$ per cent of the total accreted stellar mass. We reinforce that these phenomena are a direct result of the stellar-to-halo mass relation, which results in massive galaxies having a higher likelihood of experiencing major mergers than low mass galaxies. Our model produces a galaxy pair fraction consistent with recent observations, exhibiting a form best described by a power-law exponential function. Translating these pair fractions into merger rates results in an inaccurate prediction compared to the model intrinsic values when using published observation timescales. We find the pair fraction can be well mapped to the intrinsic merger rate by adopting an observation timescale that decreases linearly with redshift as $T_{\mathrm{obs}} = -0.36(1+z)+2.39$ [Gyr], assuming all observed pairs merge by $z=0$.
\end{abstract}




\section{Introduction}
In the hierarchical picture of galaxy formation within the $\Lambda$CDM framework, mergers play a critical role in the formation and continued evolution of galaxies. Consequently the galaxy-galaxy merger rate and its dependence on mass, mass ratio, and redshift are of fundamental interest. The frequency of galaxy mergers cannot be observed directly and so we must rely on theoretical models for galaxy formation along with a robust set of observations to ascertain the cosmological galaxy-galaxy merger rate.

Many theoretical models build upon the foundation laid by dark matter (DM) only $N$-body simulations, with each model applying a different method for populating DM haloes with their constituent galaxies. The underlying halo-halo merger has largely converged among various theoretical models \citep{Fakhouri2008, FMB10, Genel09, Genel10}. Despite this agreement in the foundational structure of galaxy evolution, theoretical models for galaxy formation have yet to establish a sufficiently accurate value for the galaxy-galaxy merger rate. There remains as a much as an order of magnitude discrepancy in the predicted values depending on mass, mass-ratio, redshift, and theoretical framework \citep{bower2006, Croton2006, Maller2006, delucia2007, font2008, Somerville2008, khochfar2009, Stewart_2009, Hopkins2010, hopkins10b, Rodriguez-Gomez2015}.

Similarly, observed merger rates have not converged so far, with different rates even derived from the same fields \citep{mantha2018, Duncan2019}. Many of the discrepancies can be attributed to varying definitions on the merger rate, including whether galaxy pairs are selected based on their stellar mass or their luminosity \citep{lotz2011, Man_2016}. Furthermore unreliable redshift measurements introduce considerable uncertainty in the selection of physically associated pairs. Additionally, merging timescales must be separately derived using theoretical models. However, considerable uncertainty remains in these merging timescales and how they might scale with redshift.

Theoretical models differ in their approach to linking dark matter haloes with galaxies. \textit{Ab initio} methods provide a complete treatment of baryonic physics to building galaxies through directly computing the physical processes. These simulations are thus reliant on accurate treatments of the physics, such as gas cooling, star formation, and the relevant feedback mechanisms \citep[]{Somerville2015, Naab2017}. Due to their sophisticated nature, they are time consuming and costly to run, which limits the resolution that can be achieved. As it is impossible to resolve the scales on which the fundamental forces act, most physical processes have to be combined into effective models -- so-called subgrid models -- with a number of free parameters, which are tuned in order to reproduce a number of observational constraints. In this sense, there are currently no true \textit{ab initio} methods in galaxy formation, as all simulations include free parameters in some form that need to be fitted or constrained by observational data, and thus rely on empirical evidence. The two most commonly used methods that aim to model the baryonic physics are \textit{hydrodynamical simulations} \citep{HorizonAGN, Hirschmann2014, Illustris_b, Illustris_a, Eagle_b, Eagle_a, TNG_a, Hopkins2018, TNG_b}, which calculate the gas physics along with the gravitational forces at the level of the resolution elements, and \textit{semi-analytical models} \citep{Kauffmann1993, Cole1994, Kauffmann1999, Somerville1999, Cole2000, Somerville2001, Baugh2006, bower2006, Somerville2008, Benson2012, Henriques2015, Somerville2015_b}, which post-process DM-only simulations and populate dark matter haloes with galaxies using analytic prescriptions at the level of individual haloes. Both approaches have made vast progress in recent years, but still struggle to reproduce a large number of observations simultaneously, as it is very difficult to explore the parameter space of the subgrid models due to the computational cost.

In this paper, we use an alternative approach known as \textit{empirical models} of galaxy formation \citep{moster2013, moster2018, moster2019, Conroy2009, Behroozi2013d, Behroozi2019}. Instead of aiming to directly model the baryonic processes, these models use parameterised relations between the properties of observed galaxies and those of simulated DM haloes. The parameters of these relations are then constrained by requiring a number of statistical observations be reproduced. This approach has the advantage of accurately matching observations by construction, allowing us to analyse the evolution of galaxy properties with cosmic time, and investigate the different growth channels. Furthermore, as these models can very efficiently post-process DM-only simulations it is easy to probe large volumes to gather statistics across a large dynamic range.

The primary goal of this paper is to determine the cosmological galaxy-galaxy merger rate, and its dependence on properties such as the stellar mass of the main galaxy, the stellar mass ratio between both galaxies, and the redshift of the merger. Instead of predicting the merger rate with a model that makes assumptions on the baryonic physics, we derive it empirically, solely based on the evolution of observables such as the stellar mass function and star formation rates, within a $\Lambda$CDM cosmology. To this end we employ the empirical galaxy formation model \textsc{emerge}\footnote{The code can be obtained at \url{https://github.com/bmoster/emerge}} \citep{moster2018, moster2019}. Additionally, we compare our results for the intrinsic merger rates, i.e. actual mergers in the model, with those produced via mock observations of galaxy pairs in order to provide a better translation between observables and underlying merger rates. We further investigate how the stellar mass of galaxies grows over cosmic time, and whether this growth mainly comes from star-formation within the galaxy, major mergers, or minor mergers. In this context, we also study how many major and minor mergers a galaxy typically has over its lifetime. We perform our analysis in the context of our empirical model, but we compare our results to observational evidence and other theoretical work to estimate the robustness of our conclusions\footnote{Scripts for reproducing our primary results can be obtained at \url{https://github.com/jaoleary}}.

This paper is organised as follows; In Section~\ref{sec:emerge} we provide an overview of the $N$-body simulation we use, as well as the empirical model used to populate the simulated dark matter haloes with galaxies. We outline our methodologies and fundamental results in Section~\ref{sec:galaxy-galaxy}, and discuss how the merger rate scales with stellar mass, mass ratio, redshift, and star formation rate. Here we also we illustrate how our model compares with other theoretical predictions. In Section~\ref{sec:obs} we discuss our results in the context of recent observations. Additionally we provide mock observations using our simulation data to create a more thorough evaluation of our model intrinsic results. Finally, in Section~\ref{sec:history} we explore the merging history of present day galaxies. Here, we determine which galaxies are grown through merging, and which type of mergers matter for stellar mass growth.

\section{Dark Matter Simulations and Emerge}
\label{sec:emerge}
Our analysis of galaxy-galaxy merger rates relies on producing galaxy merger trees encompassing a large dynamic range, occupying an appropriately large cosmic volume. We employ the empirical model \textsc{emerge} to populate dark matter haloes with galaxies based on individual halo growth histories. In this section we discuss the fundamental tools used to ultimately produce galaxy merger trees. Throughout this paper we adopt Planck $\Lambda$CDM cosmology \citep{planck} where $\Omega_m = 0.3070$, $\Omega_{\Lambda} = 0.6930$, $\Omega_b = 0.0485$, where $H_0 = 67.77\,\mathrm{km}\,\mathrm{s}^{-1}\mathrm{Mpc}^{-1}$, $n_s=0.9677$, and $\sigma_{8}=0.8149$.
\subsection{Obtaining halo merger trees}
We utilise a cosmological dark matter only $N$-body simulation in a periodic box with side lengths of $200$ Mpc. The initial conditions for this simulation were generated using \textsc{Music} \citep{music} with a power spectrum obtained from \textsc{CAMB} \citep{camb}. The simulation contains $1024^3$ dark matter particles with particle mass $2.92\times10^8\mathrm{M}_{\odot}$. The simulation was run from $z=63$ to $0$ using the Tree-PM code \textsc{Gadget3} \citep{gadget2}. In total 94 snapshots were created evenly spaced in scale factor $(\Delta a = 0.01)$. Dark matter haloes are identified in each simulation snapshot using the phase space halo finder, \textsc{Rockstar} \citep{rockstar}. Halo merger trees are constructed using \textsc{ConsistentTrees} \citep{ctrees}, providing detailed evolution of physical halo properties across time steps. Throughout this paper we use the term 'main halo' to designate haloes which do not reside within some other larger halo, and 'subhalo' to refer to haloes contained within another halo.

\subsection{Halo-Halo mergers}
\label{sec:halohalo}
Prior to evaluating the galaxy-galaxy merger rate we take a look at the halo-halo merger rate. Due to our model's reliance on the individual growth histories of dark matter haloes, it is important to verify that our simulation is assembling haloes in a manner consistent with other theoretical predictions and models \citep[][]{Genel09,Genel10,FMB10}.
We compute the halo-halo merger rate directly using the trees constructed with \textsc{ConsistentTrees}. We define halo mergers at the time when a halo first becomes a subhalo, not when the subhalo becomes distrupted. This is consistent with the definition of halo mergers adopted by \citet{Genel09,Genel10,FMB10}. The merger rate is then calculated at each redshift as a function of the descendant halo mass $M_0$, and the mass ratio $\xi = M_{i}/M_{1}$ of the progenitor haloes (for $i>1$), where $M_{1}$ is the most massive progenitor to $M_{0}$.
\begin{figure}
	\includegraphics[width=\columnwidth]{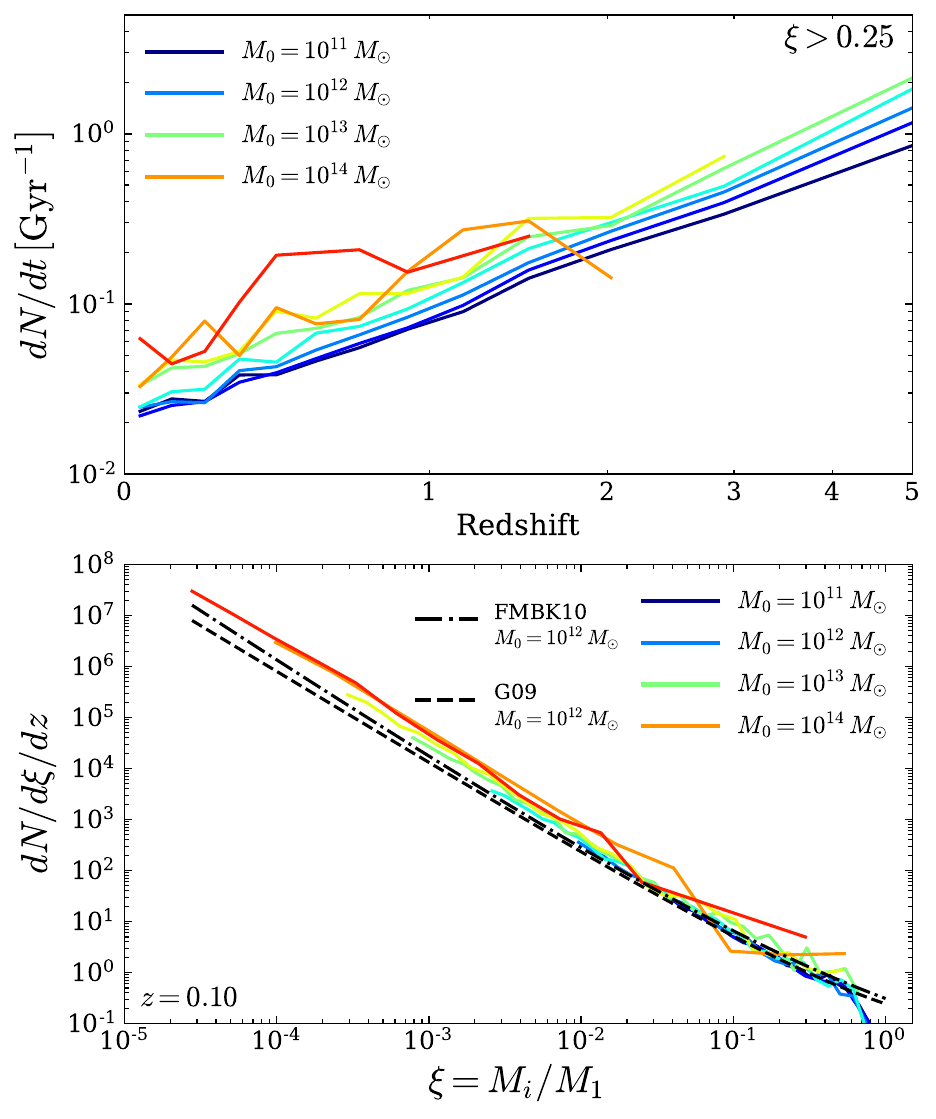}
	\caption{Halo-halo merger rates. The coloured lines indicate halo-halo merger rates from our simulation for the noted descendant halo masses. For ease of comparison we adopt the mass ratio definition of \citet{FMB10}, where $\xi=M_{i}/M_{1}=1/\mu$ (see Section~\ref{sec:galaxy-galaxy}). \textit{Top panel}: Major ($\xi>0.25$) Halo-halo merger rates \textit{per Gyr}, scaling with redshift. The bumps for higher mass haloes is due to low number statistics. \textit{Bottom panel}: Halo-halo merger rates scaling with mass ratio $\xi$ at $z=0.1$.  The black dash-dash and dash-dot lines shows the best fit merger rate for a halo with $M_0 = 10^{12}M_{\odot}$, from \citet{Genel09} and \citet{FMB10} respectively.}
	\label{fig:halohalo}
\end{figure}

Figure \ref{fig:halohalo} shows the mean halo-halo merger rate per descendant halo. When taking the halo-halo merger rate per halo, we find rates that adopt the same nearly mass-independent scaling shown in previous works \citep[][]{Genel09,Genel10,FMB10}, The top panel shows that the merger rate \textit{per Gyr} exhibits a strong power-law scaling with redshift. The bottom panel shows that, just as in previous works, we find a scaling $\propto \xi^{-2}$ for a fixed redshift interval. Our results indicate that our underlying $N$-body simulation is in agreement with other works.

At this point we are free to implement our galaxy formation model. In this process we will see how this simple universal halo merger rate becomes transformed through the complex connection between galaxies and their haloes.

\subsection{Connecting galaxies to haloes}
In the hierarchical view of galaxy formation, each galaxy starts its life at the center of an isolated halo. As the dark matter haloes grow and cannibalise one another, so too will their occupant galaxies. Empirical models populate simulated DM haloes with galaxies, and evolve each galaxy according to physically motivated parametrisations, directly constrained by real observables. Thus, these models provide a statistical link between galaxy and halo properties without the need to directly model baryonic physics. In this way, \textsc{emerge} is able to produce accurate galaxy catalogues exhibiting the range of physical properties observed in large galaxy surveys. This model additionally allows us to self-consistently track galaxies across times steps, providing the opportunity to explore and evaluate their individual growth histories.

The primary avenue for galaxy growth in \textsc{emerge} is through \textit{in-situ} star formation. Each galaxy is seeded at the center of a dark matter halo with a SFR directly driven by the growth of the dark matter halo, $\dot{M}$. On large scales, baryons are assumed to uniformly trace the underlying cosmic dark matter distribution such that each halo contains a fixed baryon fraction $f_b=\Omega_b/\Omega_m$. From this it follows that the growth rate of each halo $\dot{M}$, should be directly proportional to the rate of baryonic growth within the halo, and the SFR in the central galaxy is given by:
\begin{equation}
	\frac{\dd m_*(M, z)}{\dd t} = \frac{\dd m_{\mathrm{bary}}}{\dd t} \epsilon(M,z) = f_{\mathrm{bary}} \frac{\dd M}{\dd t}\epsilon(M,z) \;.
\end{equation}
Here, $\dot{m}_{\mathrm{bary}}(M,z)$ is the baryonic growth rate which describes how much baryonic material is becoming available, and $\epsilon(M,z)$ is the instantaneous conversion efficiency, which determines how efficiently this material can be converted into stars.

The instantaneous baryon conversion efficiency is impacted by a variety of physical processes, gas cooling, AGN feedback, supernova feedback, etc. \citep{Somerville2015, Naab2017} \textsc{emerge} seeks to establish the minimally viable parametrisation necessary to replicate observations. In the most basic picture, the instantaneous efficiency is governed only by redshift and halo mass. However, the model remains flexible as additional parameters can be added on an "as-needed" basis. In particular, it was determined that a double power-law parametrisation is sufficient to model the instantaneous baryon conversion efficiency as a function of halo mass at any redshift \citep[][]{behroozi2013, moster2018},
\begin{equation}
	\epsilon (M,z) = 2\, \epsilon_N \left[ \left(\frac{M}{M_{1}}\right)^{-\beta}+\left( \frac{M}{M_{1}}\right)^{\gamma}\right]^{-1}  \;,
	\label{eq:efficiency}
\end{equation}
where the normalisation $\epsilon_N$, the characteristic mass $M_1$, and the low and high-mass slopes $\beta$ and $\gamma$ are the free parameters used for the fitting. Furthermore, the model parameters are linearly dependent on the scale factor:
\begin{eqnarray} \label{eq:Mz}
	\log_{10} M_{1}(z) &=& M_0 + M_z\frac{z}{z+1} \;,\\
	\epsilon_N &=&  \epsilon_0 + \epsilon_z\frac{z}{z+1}\;, \\
	\beta(z) &=& \beta_0 + \beta_z\frac{z}{z+1}\;, \\
	\gamma(z) &=& \gamma_0 \;.
\end{eqnarray}
These parameters are allowed to vary freely within their boundary conditions in order to produce a fit in agreement with observation. Observables are chosen such that model parameters can be isolated and independently constrained, thus avoiding degeneracy. In particular, the characteristic mass ($M_0$ and $M_z$)is constrained by stellar mass functions (SMFs). The efficiency normalisation parameters ($\epsilon_0$ and $\epsilon_z$) can be constrained by the cosmic star formation rate density (CSFRD). The efficiency slopes ($\beta_0,\,\beta_z$ and $\gamma_0$) are constrained by specific star formation rates (sSFRs).

\subsection{Galaxy growth through mergers}
\label{sec:tdf}
Aside from \textit{in-situ} star formation, galaxy mergers are the other primary mechanism contributing to galaxy growth in \textsc{emerge}. In the context of \textsc{emerge}, we specify galaxies of three types; central, satellite and orphan. Central galaxies exist in the center of main haloes. While, satellite galaxies sit at the center of sub-haloes, orbiting within some larger main halo. Orphan galaxies were formed in the same way as satellite galaxies, however, their subhalo has since been stripped below the resolution of the halo finder. As orphans are no longer traceable in the simulation, they require special numerical treatments to address their continued evolution.

When a galaxy first becomes an orphan, a dynamical friction clock is set. We use its last known orbital parameters to compute the dynamical friction time. Specifically, we use the dynamical friction formulation specified by \citet{BK08} to control orphan orbital decay:
\begin{equation}
	t_{df}=0.0216H(z)^{-1}\frac{(M_{0}/M_{1})^2}{\ln(1+M_{0}/M_{1})}\exp(1.9\eta) \left( \frac{r_{1}}{r_{\mathrm{vir}}}\right)^{2} \;,
	\label{eq:tdf}
\end{equation}
where $H(z)$ is the Hubble parameter, $r_{\mathrm{vir}}$ is the virial radius of the main halo ($M_0$), $r_1$ is the radial position of the subhalo ($M_1$) with with respect to the center of the main halo, and $\eta$ is a measure for the orbital circularity of the subhalo.
When the dynamical friction time has elapsed the orphan galaxy will be merged with the central system where a portion of the satellite stellar mass will be added to the descendant galaxy as
\begin{equation}
	m_{\mathrm{desc}} = m_{\mathrm{main}} + m_{\mathrm{orphan}}\,(1-f_{\mathrm{esc}}),
	\label{eq:desc_mass}
\end{equation}
where $m_{\mathrm{desc}}$ is the mass of the descendant galaxy, $m_{\mathrm{main}}$ is the mass of the main progenitor galaxy, $m_{\mathrm{orphan}}$ is the mass of the progenitor orphan galaxy, and $f_{\mathrm{esc}}$ is the fraction of mass that will be distributed to the ICM during the merger. The escape fraction is a free parameter in the model and is largely constrained by the low redshift behaviour on the massive end of SMFs along with the sSFR of massive galaxies.

If however, an orphan is on its way to merge with a satellite galaxy and that target itself becomes an orphan before $t_{df}$ has elapsed, then the orphan galaxy will have its dynamical friction clock reset according to the mass of the new central system. In this special case where $t_{df}$ must be reset, we rely on a recently implemented approximation for orphan-halo mass loss.

\subsection{Orphan-halo mass loss}
\label{sec:stripping}
A prescription for orphan-halo mass loss is important for a few reasons. The first is the dependence of $t_{df}$ on the mass of both systems involved, as shown in eq.~\ref{eq:tdf}. The other reason is as a means of defining the gravitational potential of the system, which is important for galaxy stripping. When a halo has lost enough mass the galaxy at the center can also become subject to tidal forces and experiences stripping. This model implements a simple halo mass threshold below which the galaxy can no longer remain bound and will be distributed to the ICM.
\begin{equation}
	M < f_s \, M_{\mathrm{peak}}
\end{equation}
Here, $M_{\mathrm{peak}}$ is the halo peak mass, and $f_s$ is the stripping fraction.

To ensure that all galaxies, including orphans, are subject to stripping, we apply a simplified formula as a stand-in for the physical tidal stripping process experienced by sub-haloes. In this approach orphan halo mass is updated at each time step, declining at the same average rate since peak mass:

\begin{equation}
	M_{i} = M_{i-1} \left( \frac{M_{\mathrm{peak}}}{M_{\mathrm{loss}}} \right)^{-\Delta t/(t_{\mathrm{loss}}-t_{\mathrm{peak}})}
\end{equation}
Here $i$ is the index of the current snapshot and $i-1$ is the previous snapshot index, and $\Delta t$ is the amount of time elapsed between these snapshots in $\mathrm{Gyr}$. The halo is assumed to lose mass at the same average rate, with a slope defined by halo peak mass $M_\mathrm{peak}$ at $t_{\mathrm{peak}}$ and subhalo disruption mass $M_{\mathrm{loss}}$ at $t_{\mathrm{loss}}$. In the initial \textsc{emerge} release, orphans inherited their last resolved halo mass. This static halo mass made orphans impervious to stripping which leads to a more resolution dependent model. One consequence of that implementation is that a galaxy would need to be stripped \textit{prior} to entering the orphan phase, necessitating a large value for $f_s$ in order to reproduce the observed clustering values.

The stripping fraction is a free parameter in this model, and is largely driven by galaxy clustering observations. As with their halo mass, orphans also have their positions updated at each timestep until $t_{df}$ has elapsed. Orphans are placed randomly on a sphere with radius $r=r_0 f_{\mathrm{dec}}$ where $f_{\mathrm{dec}} = \sqrt{1-\Delta t/t_{df}}$, and $\Delta t$ is the time elapsed since the subhalo was last resolved \citep{BT87}. 

\begin{figure}
	\includegraphics[width=\columnwidth]{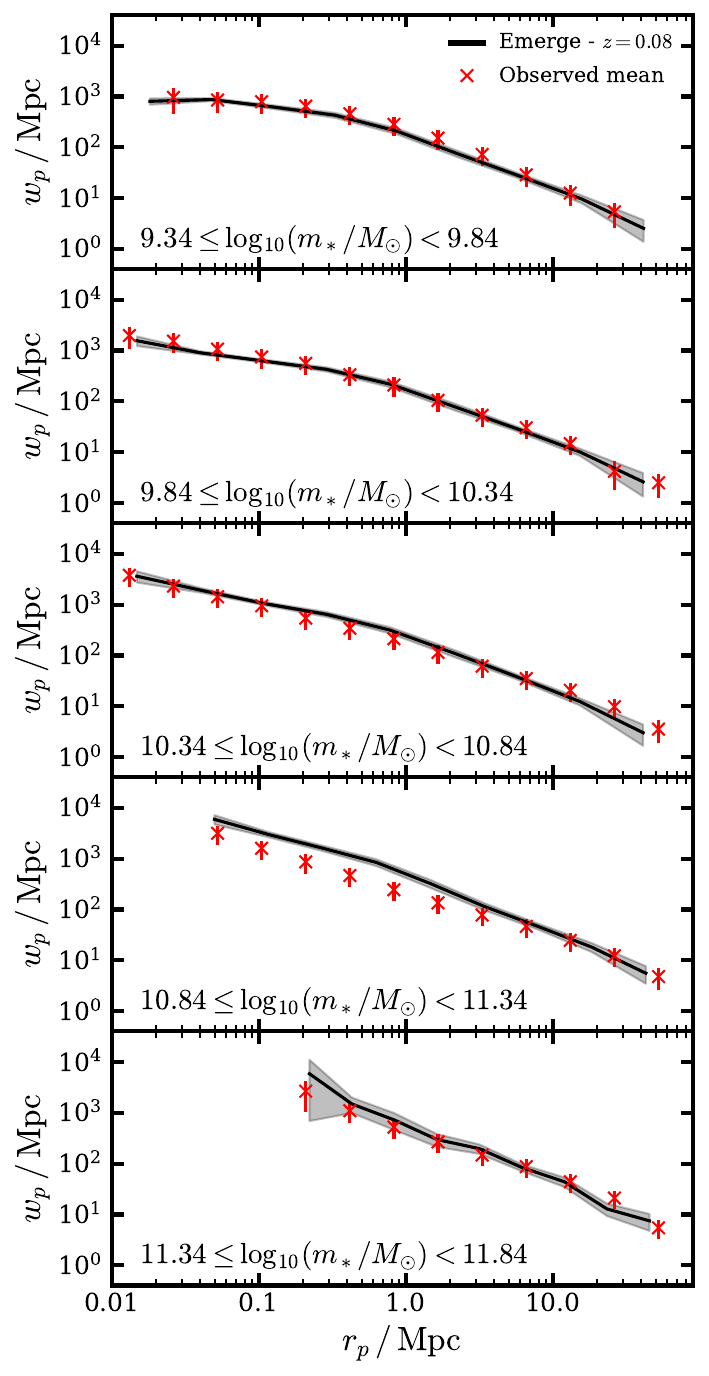}
	\caption{Clustering in the updated model down to $10$ kpc. Shaded regions illustrate the $1\sigma$ uncertainty range determined though jackknife sampling. Red crosses and associated error bars represent the mean observed $w_p$ value for each bin. The observational points represent average quantities constructed from \citet{Li2006, Guo2011} and \citet{Yang2012}.}
	\label{fig:clustering}
\end{figure}

The interplay between halo mass, stripping, and clustering make a treatment for orphan halo mass critical in fitting clustering down to $10$ kpc, which is important for determining merger rates derived through projected galaxy pairs, Section~\ref{sec:pairs}. Figure~\ref{fig:clustering} shows the projected galaxy correlation function in several mass bins under this new model variation.

When implementing these model improvements we did not alter the fitting procedure. Parameter uncertainties are determined through the same affine invariant MCMC described by \citet{Goodman2010} and employed in \citet{moster2018}. Lastly, we utilise the same observed data sets listed in \citet{moster2018}, as well as the additional data sets noted in \citet{moster2019}. The addition of this halo mass loss formulation along with new observational data naturally resulted in a set of best fit parameters different from previously published results (see Table~\ref{tab:best_fit}). While the most notable parameter change was $f_s$, other parameters have moved beyond the $1\sigma$ uncertainty ranges quoted in \citet{moster2018}. This movement does not reflect tension in the observed datasets, but is more likely the result of the complex topology of the high-dimensional parameter space. The posterior surfaces contains local minima separated by many sigma from one another with each exhibiting a reasonable fit to the observations.

\begin{table}
	\centering
	\caption{The best fit model parameters used for this work.}
	\label{tab:best_fit}
	\begin{tabular}{lccr} 
		\hline
		Parameter    & Best-fit  & Upper $1\sigma$ & lower $1\sigma$ \\
		\hline
		\hline
		$M_0$        & 11.34829  & +0.03925        & -0.04153        \\
		$M_z$        & 0.654238  & +0.08005        & -0.07242        \\
		$\epsilon_0$ & 0.009010  & +0.00657        & -0.00451        \\
		$\epsilon_z$ & 0.596666  & +0.02880        & -0.02366        \\
		$\beta_0$    & 3.094621  & +0.15251        & -0.14964        \\
		$\beta_z$    & -2.019841 & +0.22206        & -0.20921        \\
		$\gamma_0$   & 1.107304  & +0.05880        & -0.05280        \\
		\hline
		$f_{esc}$    & 0.562183  & +0.02840        & -0.03160        \\
		$f_{s}$      & 0.004015  & +0.00209        & -0.00141        \\
		$\tau_{0}$   & 4.461039  & +0.42511        & -0.40187        \\
		$\tau_{s}$   & 0.346817  & +0.04501        & -0.04265        \\
		\hline
	\end{tabular}
\end{table}

\subsection{Satellite quenching}
Galaxy quenching is one other mechanism that affects the growth of galaxies. If a dark matter halo begins to become accreted by a larger halo, its own growth rate will decline. At some point the halo will reach its peak mass $M_{\mathrm{peak}}$ after which the halo will not grow, consequently reducing the `inflow' of gas. After some time the galaxy at the center of such a halo will deplete the remaining cold gas supply through star formation and become quenched.

To address star formation in these galaxies \textsc{emerge} invokes a `delayed-then-rapid' model for quenching \citep{Wetzel2013}. In this model, after a halo has reached peak mass the central galaxy will continue to form stars at a constant rate equal to the star formation rate at $t_{\mathrm{peak}}$. After a time $\tau$ the cold gas supply is assumed depleted and the star formation rate will be set to $0$. The quenching timescale can be parameterised as:
\begin{equation}
	\tau = t_{\mathrm{dyn}} \, \tau_{0} \left( \frac{m_{*}}{10^{10}M_{\odot}}\right)^{-\tau_{\mathrm{s}}} \;.
\end{equation}
Here $t_{\mathrm{dyn}}$ is the halo's dynamical time and $\tau_{0}$ is normalization, which together specifying a minimum quenching time of $t_{\mathrm{dyn}} \, \tau_{0}$. The normalisation determines the quenching timescale for galaxies with $m_*\geq10^{10}M_{\odot}$ while the slope $\tau_\mathrm{s}$ describes the quenching timescale for low mass galaxies. These parameters are largely constrained by the observed fraction of quenched galaxies at several redshifts.

\section{The Galaxy-Galaxy merger rate}
\label{sec:galaxy-galaxy}
In this section we discuss the galaxy-galaxy merger rates intrinsic to and derived from \textsc{emerge}. First we present the \textit{intrinsic} merger rate, that is the rate at which galaxies are merged using the processes outlined in Section~\ref{sec:tdf}. The intrinsic merger rate provides insight into the actual buildup of stellar material using the internal mechanics of the empirical model. We then present a merger rate derived using mock observations applied to mock galaxy catalogues. This provides a bridge to more completely address any discrepancies between theoretical models and observations.

First we should address some terminology common to both approaches. Each galaxy merger can be classified in terms of stellar mass and stellar mass ratio:
\begin{description}
	\item $m_0$: The stellar mass of the \textit{descendant} galaxy at the snapshot following the merger.
	\item $m_1$: The stellar mass of the \textit{main} progenitor galaxy defined at the snapshot just prior to the merger.
	\item $m_2$: The stellar mass of the co-progenitor galaxy at the snapshot just prior to the merger.
	\item $\mu$: The stellar mass ratio taken with respect to the progenitor galaxies, $\mu \equiv m_1/m_2$. In the most general case the main progenitor $m_1$ is \textit{also} the most massive progenitor. Due to scatter in the stellar-to-halo mass relation (SHMR) there are some scenarios under which $m_2$ > $m_1$. In these cases we invert this relation such that $\mu \geq 1$. These special cases represent fewer than $5$ per cent of all mergers with a descendant mass larger than $10^{9} M_{\odot}$.
\end{description}
\subsection{Intrinsic merger rate}
\label{sec:intrinsic}
Having constructed galaxy merger trees, computing the merger rate is straight forward. In the trees we identify galaxy mergers as any pair of galaxies sharing an identical descendant galaxy. In each case we assume every merger is binary and occurs instantaneously at $t_{df}$.\footnote{This binary assumption holds true for $> 98$ per cent of all mergers with a descendant mass $>10^{9} M_{\odot}$. Despite this small number of cases we nonetheless adjust the descendant mass of each merger to ensure the descendant mass only reflects mass contributed by the most massive progenitor and the merging satellite.} We provide two measures for the merger rate: merger rate per comoving volume and merger rate per galaxy. The process is similar in each case, with the difference arising in how the rate is normalised. The first step is to construct bins for time, stellar mass and mass ratio. For each time bin we count the number of mergers which satisfy the mass and mass ratio requirements, and whose merging time ($t_{df}$) resides within that bin. When computing the merger rate per comoving volume we then divide the number of mergers $(N_{\mathrm{\mathrm{merge}}})$ in each bin by the bin widths $\dd t$ and by the volume of our box, so the merger rate is given by:
\begin{equation}
	\Gamma = \frac{N_{\mathrm{merge}}}{\dd V\dd t} \left[\mathrm{cMpc}^{-3} \mathrm{Gyr}^{-1}\right] \;.
\end{equation}
For the merger rate per galaxy we instead divide the number of mergers in each bin by the bin width, and the number of galaxies $(N_{\mathrm{gal}})$ at the central redshift of the bin. At these redshifts $N_{\mathrm{gal}}$ is determined by linearly interpolating counts between the nearest two snapshots. The merger rate per galaxy is then described by:
\begin{equation}
	\mathfrak{R} = \frac{N_{\mathrm{merge}}}{N_{\mathrm{gal}}\,\dd t} \left[\mathrm{Gyr}^{-1}\right]
	\label{eq:rate_per_gal}
\end{equation}

While operating on the same data the two measures for merger rate produce qualitative differences due to the scaling of the merger rate per galaxy with the number density of galaxies. For this reason the merger rate per galaxy is often the preferred measure as it is more robust against cosmic variance. We explore both rate measures to provide a more complete comparison with other works.

\subsubsection{Scaling with mass and redshift}
\begin{figure*}
	\includegraphics[width=\textwidth]{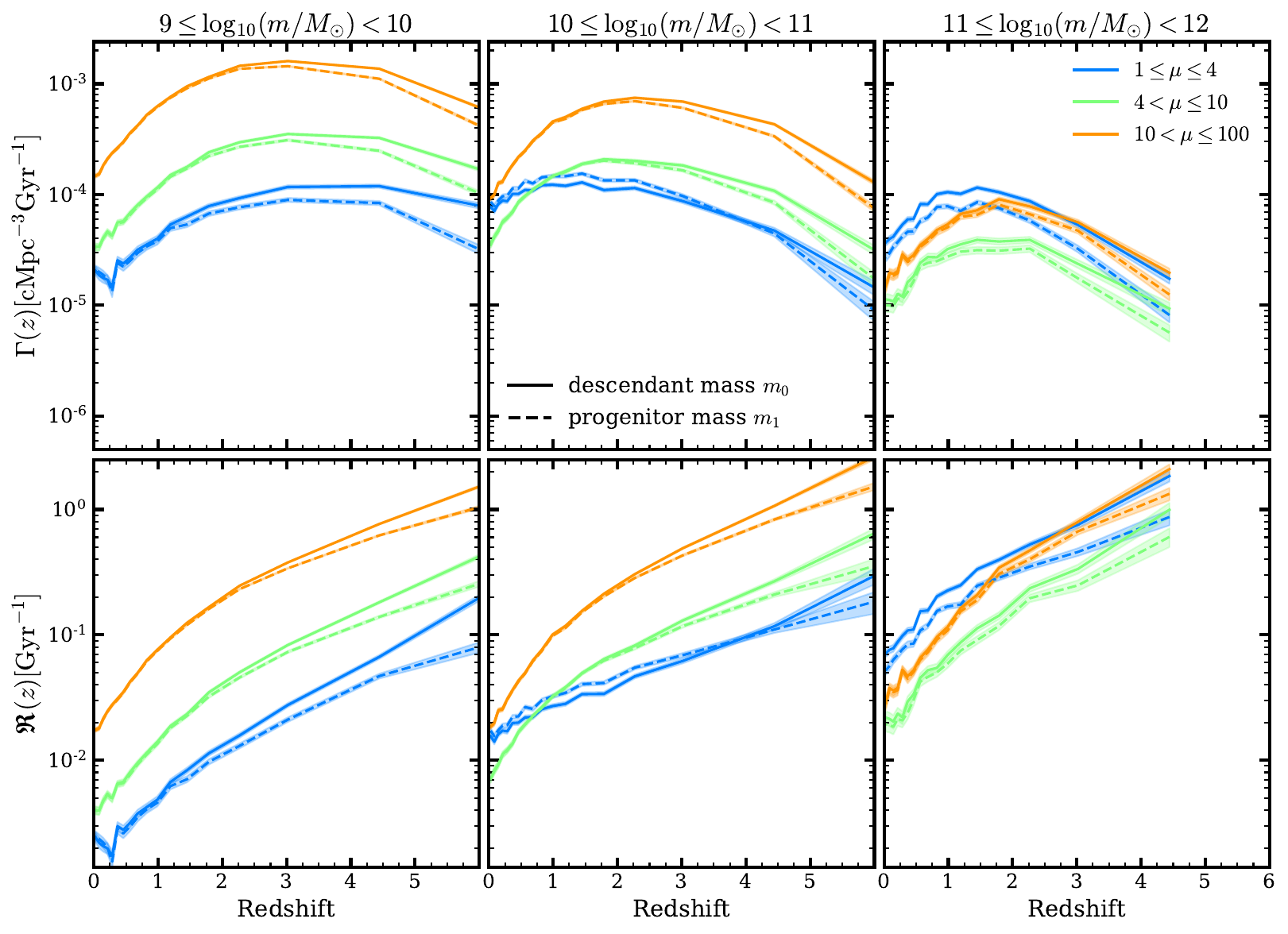}
	\caption{The galaxy-galaxy merger rates as a function of redshift and mass ratio, $\mu \equiv m_1 / m_2$. The top row of panels show the merger rate density $\Gamma(z)$, i.e. the total number of mergers per \textit{comoving} volume, and the lower panels show the merger rate \textit{per galaxy} $\mathfrak{R}(z)$. Solid lines indicate galaxy mergers selected based on the descendant mass of the merging system, $m_{0}$. In this scenario the progenitor galaxies $(m_{1}, \,m_{2})$ are permitted to have masses outside the noted mass bands. The dashed lines show mergers selected based on the \textit{main} progenitor mass $(m_1)$ of the merging systems. Here only the main progenitor must reside in the specified mass band. The shaded regions indicate Poisson error in the number of mergers.}
	\label{fig:rate_vs_redshift}
\end{figure*}

Figure~\ref{fig:rate_vs_redshift} illustrates the merger rate scaling with redshift for three different mass bins. For each mass bin we select mergers based on descendant mass $m_0$ (solid lines), or  main progenitor mass $m_1$ (dashed lines). Selection based on descendant mass probes galaxy mergers after their mass contributions have been added into the descendant galaxy. This is a common approached taken by other theoretical works exploring galaxy merger rates. This method presumes we have knowledge on how the mass of two galaxies is added together in a merger, something not easily accessible to observations. Selecting mergers based on progenitor properties provides a more observer friendly measure of the merger rate as it can be more directly compared with observationally derived rates (see Section \ref{sec:pairs}).

In each mass band we show the merger rate for three different mass ratio intervals. The first, and arguably the most important, are called \textit{major mergers} (blue lines), although the precise mass ratio that defines a major merger is not well defined and varies across literature. The key characteristic of major mergers are their transformative properties. Such mergers are very disruptive to both systems, suspected of prompting drastic changes in stellar populations and descendant morphologies. In this paper we take major mergers to be $1 \leq \mu \leq 4$. The next mass ratio interval are similarly labelled \textit{minor mergers} (green lines). Minor mergers while not as individually disruptive as their larger counterparts, still contribute to the evolutionary process of large galaxies. There is evidence to suggest that such mergers, if occurring at high enough frequency, can produce some of the same morphological changes generated through major mergers \citep[]{Naab2009, Oser2010, Hilz2012, Hilz2013, Karademir2019}, lead to the thickening of disc galaxies \citep{Abadi2003, Kazantzidis2008, Purcell2009, Moster2010b, moster2013}, and even drive the rotation speed of massive early type galaxies \citep[]{Bois2011, Moody2014, Penoyre2017, Schulze2018} . The final category are so called \textit{mini mergers} (orange lines). As their name entails, this category represents the smallest merging events. While not terribly transformative, understanding their frequency is helpful for constructing a complete mass budget and internal radial distribution for the accreted material of a galaxy.

The merger rate per comoving volume ($\Gamma$) exhibits a nearly mass independent shape in the number of mergers occurring. For each mass and mass ratio interval we find a sharp increase in the number of mergers at low redshift, with a well defined peak $1 \lesssim z\lesssim 2$. Beyond the peak we see a rapid decay in the total number of mergers towards high redshift. When we instead take the merger rate in the context of an evolving galaxy population ($\mathfrak{R_{\mathrm{merge}}}$) we find quite a different trend. In general we find the merger rate increases with redshift at all masses and mass ratios. However, our results do not show a simple power-law scaling with redshift. For major mergers, at low and intermediate mass, we find an excess over a power-law for $z\gtrsim3$. This break from a power-law redshift scaling is primarily evident when selecting mergers based on descendant galaxy stellar mass.

Additionally, We can observe a discrepancy in the merger rate when selecting merger rates based on progenitor mass vs. descendant mass. In the lowest mass bins the difference in these two quantities only becomes manifest at higher redshifts ($z\gtrsim2$), where selecting based on progenitor mass produces a noticeably lower merger rate. However, for the most massive galaxies the difference is more dramatic. We find that in this mass range both measures produce functionally similar results, with a nearly constant scaling offset, where the descendant mass selected major merger rate is a factor of $1.5-2$ larger than that produced with progenitor selection. To explain this, we should recall that in our approach the average number of galaxies remains fixed for any time and mass interval, that is $N_{\mathrm{gal}}$ of eq.~\ref{eq:rate_per_gal} remains the same regardless of whether we compute the rate based on descendant mass or progenitor mass. Additionally, for a lower mass threshold, there are in general more galaxies that fit within our mass range at timestep $i$ compared to $i-1$. At the highest masses and redshifts this effect is amplified due to the low numbers of galaxies present. This is similar to the argument by \citet[]{Genel09} to explain such differences in the context of halo-halo merger rates.

Lastly, we can look across the mass panels to see how the frequency of major mergers changes with redshift. For low-mass galaxies it is clear that minor and mini mergers dominate. When moving to intermediate-mass ranges, galaxies like the Milky Way, we find that for the first half of cosmic time minor and mini mergers occur at greater frequency than major mergers. Near $z\approx1$ this changes, as major mergers become more frequent than minor mergers and occur at nearly the same frequency as mini mergers by $z=0$. Finally, for the most massive galaxies, at $z\lesssim3$ major mergers are the most frequent, with mini mergers being the next most common, and minor mergers making up the smallest fraction of mergers. We explore these strong mass ratio dependencies of the merger rate in closer detail in the next section.

\subsubsection{Scaling with mass ratio}
\label{sec:mass_ratio}
\begin{figure*}
	\includegraphics[width=\textwidth]{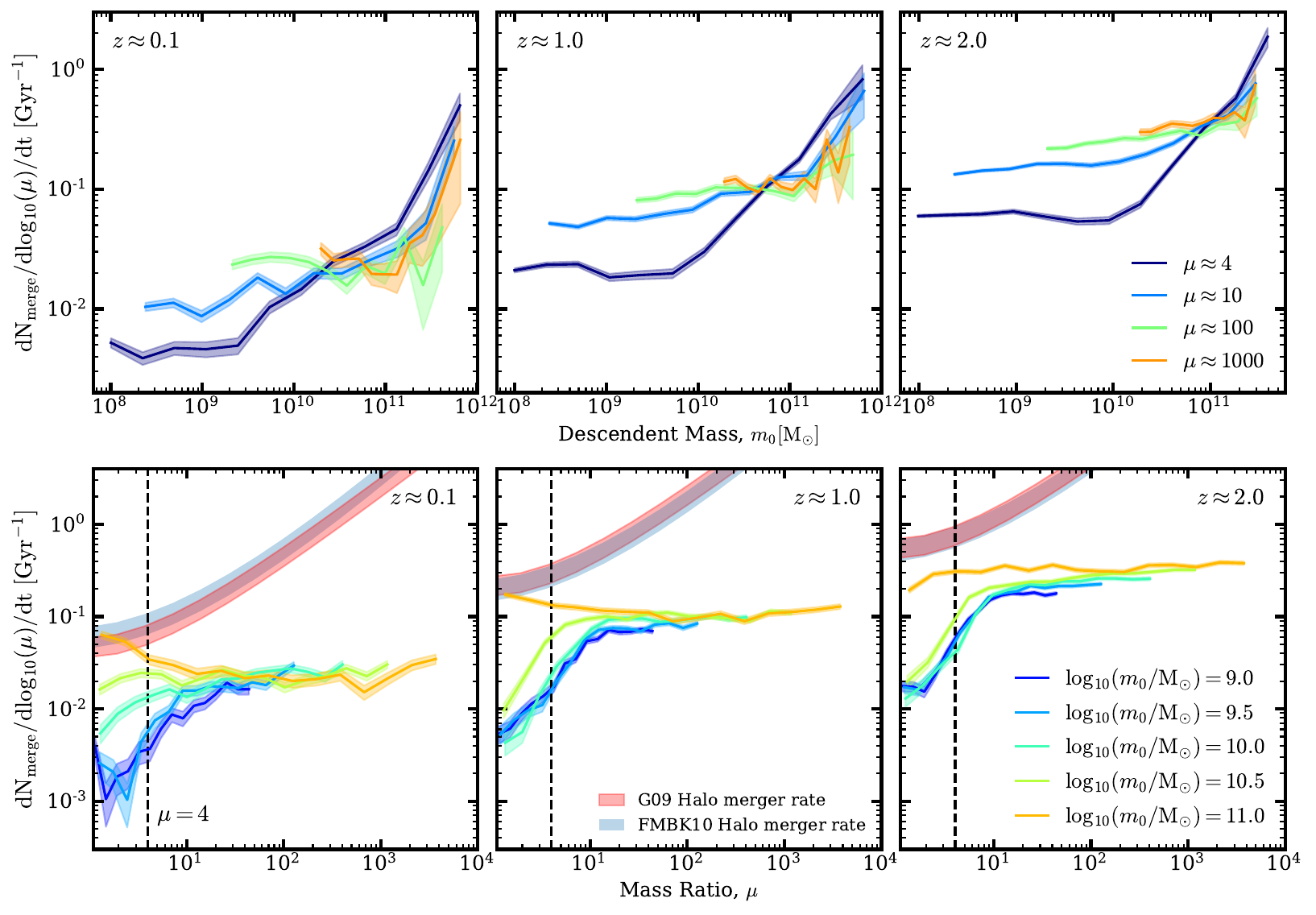}
	\caption{The galaxy-galaxy merger rate as a function of descendant stellar mass and progenitor mass ratio. \textit{Top panel}: The merger rate per galaxy as a function of descendant stellar mass, $m_{0}$. The coloured lines specify the merger rate for a specific mass ratio $\mu$. Shaded regions indicate Poisson error in the number of mergers. \textit{Bottom panel}: The merger rate per galaxy as a function of progenitor mass ratio. Each line includes mergers with a descendant mass noted by the colour. The x-axis shows the distribution of merger mass ratios experienced within each mass band. The vertical dashed black line denotes the threshold for major mergers at $\mu = 4$. Shaded regions surrounding solid lines indicate Poisson error in the number of mergers. The red and blue shaded regions show the best fit halo-halo merger rate for haloes with $11.25 \leq \log_{10}(M/M_{\odot}) < 13$, from \citet[]{Genel09} and \citet[]{FMB10} respectively.}
	\label{fig:mass_vs_ratio}
\end{figure*}
\begin{figure}
	\includegraphics[width=\columnwidth]{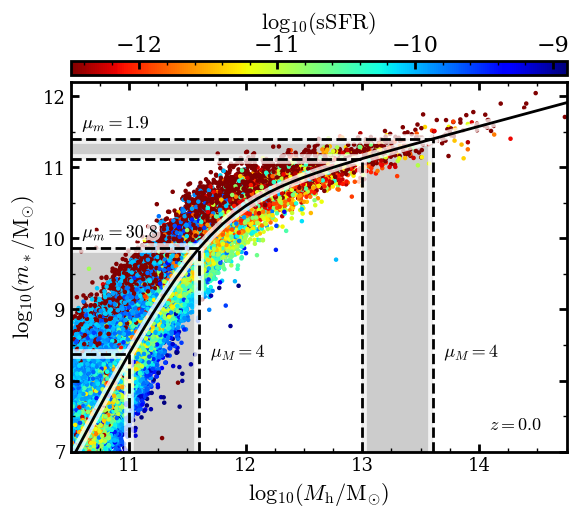}
	\caption{The $z=0$ stellar-to-halo mass relation produced by \textsc{emerge}. The solid line illustrates the best fit over the mean values of the data. The shaded regions show how a major merger in halo mass $(\mu_{M}=4)$ translates to galaxy-stellar mass ratio $(\mu_{m})$ on both the low and high-mass end of the stellar to halo mass relation. Along the best fit curve we can see that a major halo-halo merger on the low-mass end will result in a very minor merger in stellar mass. On the high-mass end a major halo-halo merger will result in a major merger in stellar mass.}
	\label{fig:shm}
\end{figure}
\begin{figure*}
	\includegraphics[width=\textwidth]{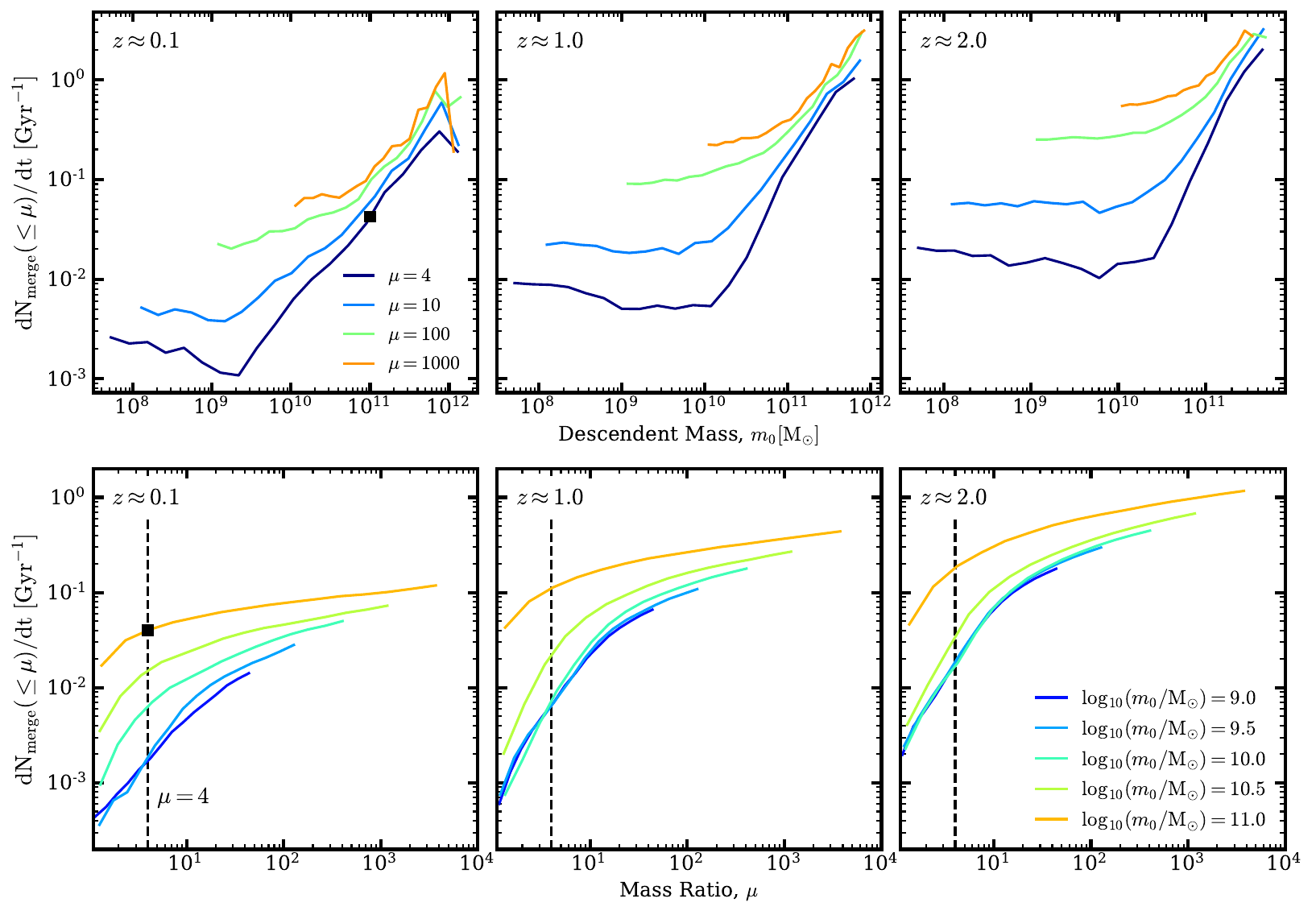}
	\caption{The \textit{cumulative} galaxy-galaxy merger rate as a function of descendant stellar mass and progenitor mass ratio. The data shown here is the cumulative histogram of the data displayed in Fig. \ref{fig:mass_vs_ratio}. The cumulative merger rate can be used to determine the merger rate for a desired mass and mass ratio. \textit{Top panel}: The cumulative merger rate per galaxy as a function of descendant stellar mass, $m_{0}$. The coloured lines each specify a different mass ratio threshold, and include all mergers below that threshold. \textit{Bottom panel}: The merger rate per mass ratio interval. Each line includes mergers with a descendant mass noted by the colour. The x-axis therefore shows the distribution of merger mass ratios experienced within each mass band. The black line denotes the threshold for major mergers $\mu=4$. As an example (square black point) we can see that at $z=0.1$ for $\log_{10}(m_0/M_{\odot})=11.0$ approximately $4$ per cent of galaxies will experience a merger in the next Gyr.}
	\label{fig:cumulative}
\end{figure*}

In this section we address the galaxy-galaxy merger rate scaling with mass ratio. We have already seen in the previous section that the merger rate density and rate per galaxy appear to possess a strong scaling with mass for a fixed mass ratio interval. Most notably, major mergers dominate at high-mass, and mini/minor mergers dominate at low masses. Our goal here is to investigate this inflection with a finer $\mu$ binning in order to explain this phenomenon.

Figure \ref{fig:mass_vs_ratio} explores the relationship between descendant stellar mass and merger mass ratio. To provide a more granular view of the mass ratio distribution, without influence from the explicit definition of major(minor) merger, we express these results as the merger rate per galaxy \textit{per $\log_{10}$ mass ratio interval} $N_{\mathrm{merge}}/\dd\log_{10}(\mu)/\dd t$. In the top panel we take mergers of a fixed mass ratio at a fixed redshift. We then bin those galaxies according to the stellar mass of their descendant system. This way we are able to see the relative importance of some merger as a function of mass. We selected four target mass ratios with a uniform log space bin of $0.45$ dex. In general each mass ratio shows a similar qualitative trend. At lower masses the slope stays relatively flat, approaching high masses each curve show an inflection point when more massive galaxies begin to experience more mergers. Looking closely we see that the inflection occurs at lower stellar mass for lower redshifts.

Similarly, The bottom panel of Figure~\ref{fig:mass_vs_ratio} shows how merger mass ratios are distributed for a fixed descendant mass and redshift. This plot is analogous to Figure~\ref{fig:halohalo} for the halo-halo merger rate. In this comparison we can immediately note some glaring differences compared to the halo-halo merger rate, most notably the galaxy-galaxy merger rate does \textit{not} show the same mass independent merger rate with respect to mass ratio. In the case of galaxy-galaxy mergers we break from a simple power law to a more complex relation with a clear redshift, mass ratio and strong mass dependence. For mergers with $\mu\gtrsim10$ all masses exhibit a similar merger rate, with some minor scaling differences with increasing redshift. Below $\mu\approx10$ we find a greater dependence on mass. In this regime the most massive galaxies maintain a nearly constant slope for all $\mu$, even experiencing a boost in the major merger rate.. Conversely, for low-mass and intermediate-mass systems we see a dropping merger rate, illustrating suppressed major mergers. Narrowing in on galaxies with $\log_{10}(m_0/M_{\odot})=10.5$, we see an interesting evolution. At $z\approx2$ these galaxies scale with mass ratio similar to the lowest mass galaxies, with a suppressed rate for $\mu \lesssim 10$. As we transition to lower redshift we can see this relationship change, with major mergers becoming more prevalent at lower redshift, and the once decreasing trend bending up to meet the clean scaling seen for massive galaxies. Both of these trends can be explained in the context of an evolving stellar-to-halo mass relation (SHMR).

First, why are major mergers suppressed for low-mass galaxies? The first driver for this effect can be seen directly in the SHMR. Figure~\ref{fig:shm} shows the SHMR present in our model for galaxies at $z=0$. Once again recalling the simple relation assumed by the halo-halo merger rate (Figure~\ref{fig:halohalo}), we can trace how a major merger in halo mass would translate to mergers in galaxy stellar mass. In this thought experiment we can make the presumption that any halo-halo merger will eventually result in a galaxy-galaxy merger. Starting along the low-mass slope of the SHMR, we note that if we select the average galaxy masses for a fixed halo mass, a major merger in halo mass would translate to a mini-merger ($\mu\approx30$) in galaxy stellar mass. Conversely, we can observe the opposite scenario on the high-mass slope, beyond the turnover. Due to the shallow slope in the SHMR for high masses we can see that a major halo-halo merger has a much larger likelihood of also leading to a major merger in galaxy stellar mass. In short, if the slope of the SHMR is unity, we would expect a major halo-halo merger to directly lead to a major galaxy-galaxy merger. Subsequently, where the slope is greater than unity we expect a suppression of major mergers, and where less than unity we expect an enhanced rate of major mergers. The second driving factor is dynamical friction. We see that small satellites orbiting a massive central galaxies have much longer dynamical frictions times (eq.~\ref{eq:tdf}) and would simply not have had enough time to merge. This effect prevents minor halo mergers from being transformed into major mergers where the stellar mass ratio of the two systems would otherwise be sufficient.

Now, why do we see that intermediate-mass galaxies with $\log_{10}(m_0/M_{\odot}) \approx 10.5 $ have suppressed major mergers at high redshift but not at low redshift? Here we move beyond the static low redshift SHMR, and instead focus on how the SHMR evolves. These intermediate-mass galaxies reside close to the turnover on the SHMR. From $z=2$ to $z=0$ we see the average halo mass for such galaxies increase by $\sim 0.08$ dex. That is these galaxies tend to live in larger haloes at lower redshift. Additionally, the turnover in the SHMR has a mild shift to lower halo mass by $~\sim 0.17$ dex. These combined effects mean these galaxies tend to sit higher along the turnover where the slope approaches unity. Thus, these galaxies begin to experience more major mergers with decreasing redshift. This can be clearly seen in the much flatter major merger rate scaling with redshift shown in Figure~\ref{fig:rate_vs_redshift}. The suppression and enhancement of major galaxy mergers as a consequence of the SHMR has been explored in the context of other semi-empirical models \citep{Stewart_2009, Hopkins2010}. However, it is important to stress the necessity of a model that can accurately reproduce the observed data (e.g. SMFs, cosmic and specific SFRs). Models that show significant deviations from these fundamental observations, or cannot self-consistently track their redshift evolution may arrive at differing conclusions regarding galaxy merger rates or galaxy mass assembly.

Finally, in Figure~\ref{fig:cumulative} we provide the cumulative galaxy-galaxy merger rates with respect to mass ratio. The information contained in the cumulative merger rates is identical to that of Figure~\ref{fig:mass_vs_ratio}. Absent a generalised fitting function for our results, the cumulative rates provide a quicker reference for determining the number of mergers occurring at some descendant mass for a given mass ratio interval.

\subsubsection{Active vs. Passive galaxies}
\label{sec:red_vs_blue}
\begin{figure*}
	\includegraphics[width=\textwidth]{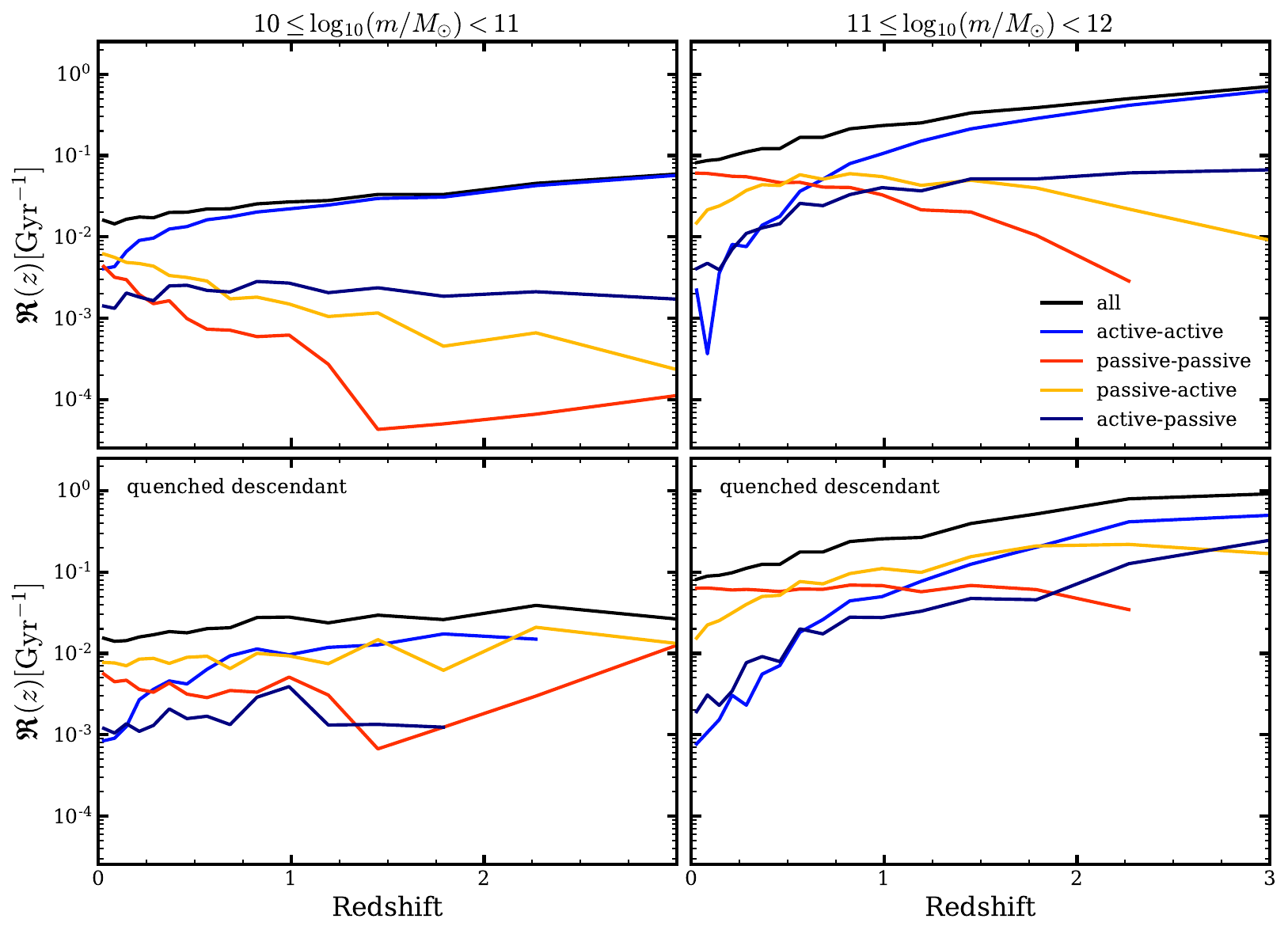}
	\caption{The major merger rate $(\mu \leq 4)$ as a function redshift for various active/passive progenitor configurations. We show the four possible scenarios for active/passive combinations in progenitor systems. Active-active would specify a merger where both progenitor galaxies are actively star forming, passive-active specifies a configuration where the main progenitor is passive (quenched) and the secondary progenitor is active (star forming), etc. \textit{Top Panel}:  The galaxy merger rate dependency on star forming properties of the progenitor galaxies. The black line shows the total merger rate for all star forming configurations of the descendant system. \textit{Bottom Panel}: The galaxy merger rates only considering mergers with a quenched descendant galaxy. At high redshift most mergers occur between active progenitors. There is a transition near $z\approx0.5$ where mergers start to become dominated by passive progenitors.}
	\label{fig:red_vs_blue}
\end{figure*}

Finally we address SFR dependencies of the cosmic galaxy-galaxy merger rate. Once again looking back to the baryon conversion efficiency (eq.~\ref{eq:efficiency}) of galaxies we can see a characteristic halo mass (see Table~\ref{tab:best_fit} and eq.~\ref{eq:Mz}) at which a galaxy is most efficient at converting gas into stars. One conclusion from this relation is that larger galaxies are inefficient at creating new stars. Thus it is important to understand the merger rate for these specific galaxies to learn how galaxy mergers drive their galaxies' continued mass growth \citep[]{khochfar2009}. Specifically we would like to know if these galaxies are grown through the merging of other large quenched galaxies, or constructed more slowly through the accretion of smaller star forming satellites. Further, understanding these mergers may help explain how mergers initiate star formation, or power AGN \citep[]{Hirschmann2010, Hirschmann2014, Hirschmann2017, TNG_a, Choi2018, Steinborn2018}.

We begin by defining our galaxies in terms of their star formation properties. Broadly this means designating a galaxy as \textit{passive} or \textit{active}, where passive galaxies are quenched and active galaxies are actively star forming. We adopt the quenching criteria of \citet{SFR} to make this distinction, where a galaxy is considered quenched if:
\begin{equation}
	\Psi < 0.3t_{z}^{-1} \;,
\end{equation}
where $t_{z}$ is the age of the universe at redshift $z$, and $\Psi$ is the specific star formation rate.

In Figure~\ref{fig:red_vs_blue} we illustrate how the \textit{major} merger rate scales when selecting mergers based on the star formation properties of their progenitor and descendant systems. We perform this in two mass bands for galaxies with $\log_{10}(m)/M_{\odot}\geq10$. For each mass bin we compute a global merger rate, only considering the star formation of the progenitor galaxies. Additionally, we perform the same analysis only considering mergers with a quenched descendant galaxy. We designate four different scenarios based on the star forming properties of the progenitor galaxies:
\begin{description}
	\item Active-active: Both progenitors are active.
	\item Passive-passive: Both progenitors are passive.
	\item Passive-active: The main progenitor is passive and the secondary progenitor is active.
	\item Active-passive: The main progenitor is active and the secondary progenitor is passive.
\end{description}

In Figure~\ref{fig:red_vs_blue} we compare the total merger rates when considering all galaxies (upper panels) versus only considering mergers with a quenched descendant galaxy (lower panels). For the redshift range shown we find very little difference in the total merger rates. This suggests that most major mergers are occurring in dense environments around an already quenched central galaxy. For the most massive galaxies, by $z\approx1$ most mergers are occurring between two passive galaxies (red lines), or between a passive central galaxy and an active satellite (yellow lines). Beyond $z\approx1$ most mergers are occurring between active galaxies (blue lines). When considering only mergers with a quenched descendant (bottom panel) we find a nearly constant merger rate if the major galaxy in the merger is already quenched. There are several different effects that lead to this result. The first is the prevalence of gas-rich (active) galaxies at high redshift making the likelihood of active-active mergers greater. The second being that if a central galaxy is quenched, it is likely that its descendant galaxy will also be quenched. Due to the lack of passive galaxies at high $z$ these results are more uncertain than the results shown in Figure~\ref{fig:rate_vs_redshift}. For the lower mass bin, mergers involving any quenched component cannot be constrained to better than a factor $\sim 2-3$ for $z\gtrsim1.5$. For the higher mass bin we see a similar degree of uncertainty for $z\gtrsim 2.25$

\subsection{Comparison to other theoretical predictions}
\label{sec:models}
\begin{figure*}
	\includegraphics[width=\textwidth]{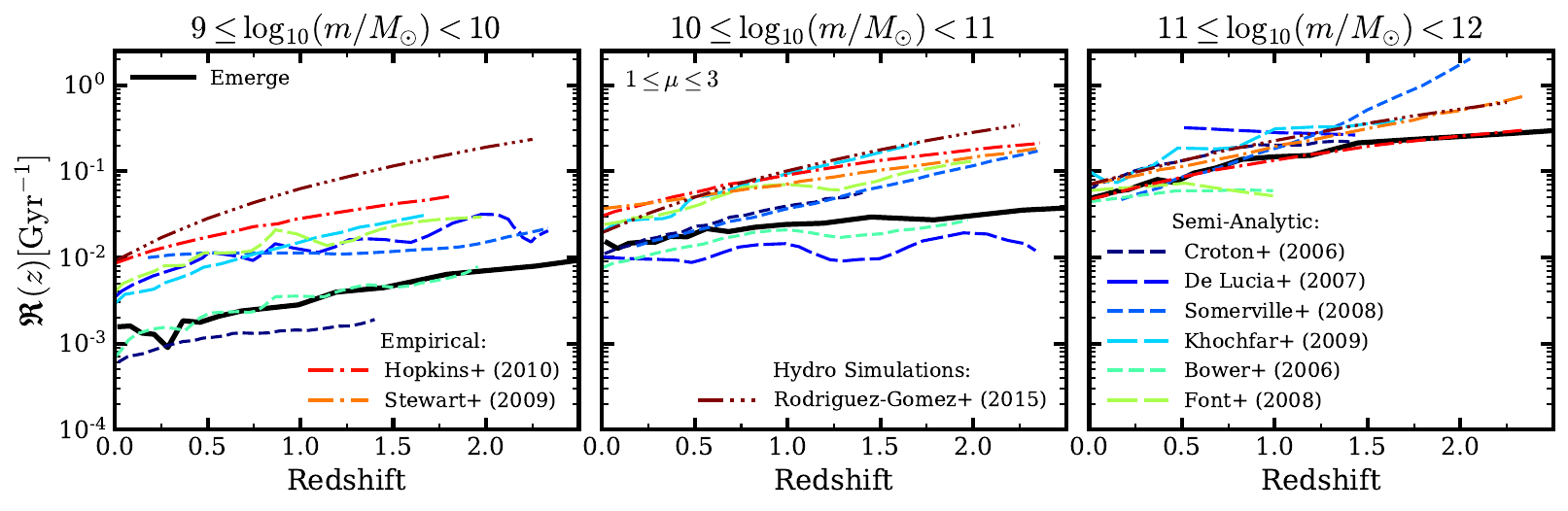}
	\caption{A comparison of merger rate per galaxy $(\mu \leq 3)$ for other theoretical models, adapted from \citet{hopkins10b}. \textit{Semi-analytic}: \citet{Croton2006, delucia2007, Somerville2008, khochfar2009, bower2006, font2008}. \textit{Empirical}: \citet{Hopkins2010, Stewart_2009}. \textit{Hydrodynamical simulations}: \citet{Rodriguez-Gomez2015}. In general our model produces a rate lower than most other models, and is in agreement with the results of \citet{bower2006}.}
	\label{fig:models}
\end{figure*}

Although most theoretical predictions are based on the same $\Lambda$CDM framework, the methods used to link DM haloes and galaxy properties has direct consequences on the predicted merger rates. Figure~\ref{fig:models} displays a side-by-side comparison of galaxy merger rates in three different mass bands produced by a diverse set of models. While our results might initially be surprising, we can see that within the context of other theoretical predictions we are firmly within a previously established range of merger rates. 

Theoretical methods for determining the merger rate can be roughly broken down into a few categories:
\begin{enumerate}
    \item Halo-halo: Assume an average halo-halo merger rate with a dynamical friction delay set at $R_{\mathrm{vir}}$. Haloes are populated with galaxies according to a SHMR \citep[e.g.][]{Hopkins2010}. 
    \item Subhalo disruption: Subhalo disruption rates are convolved with some SHMR with no delay applied \citep[e.g.][]{Stewart_2009, Rodriguez-Puebla2017}.
    \item Halo merger trees without substructure: Halo merger trees (EPS or $N$-body) are populated with galaxies according to a model. A dynamical friction delay is applied to satellites at $R_{\mathrm{vir}}$ \citep[][]{Somerville2008, khochfar2009, font2008}.
    \item Halo merger trees with substructure: Sub-haloes are tracked within an $N$-body simulation, where galaxies are populated according to a model. A dynamical friction delay is applied to satellite galaxies when their $N$-body subhalo is disrupted \citep[e.g. this work, ][]{bower2006, Croton2006, delucia2007}.
    \item Hydro: The baryonic components of galaxies in hydrodynamical simulations are tracked to determine when they coalesce \citep[e.g.][]{Rodriguez-Gomez2015}.
\end{enumerate}

At low and intermediate masses our model predicts a merger rate as much as an order of magnitude lower than some other predictions. Additionally, due to the more shallow scaling with redshift, our results deviate from other models more strongly at higher redshifts, particularly at intermediate masses. We do, however, find our results to be in good agreement with those of \citet[][]{bower2006}, who employ a semi-analytic model on top of $N$-body merger trees. At the highest masses models tend to agree more closely in terms of magnitude and redshift scaling of the merger rate. 

The categories noted above broadly describe the merging process in each modelling strategy, but do not classify all model options that could impact the merger rate. In particular the calibrated or predicted SMF, as well as the chosen treatments for orphan/satellite stripping could play a factor in the resulting merger rates within each model. Any model implementations that could impact the SHMR or cause premature/prolonged merging would also impact the rates, regardless of the guiding dark matter component of the model. A more complete overview of the models shown in Figure~\ref{fig:models}, as well as the systematic uncertainties in determining merger rates can be found in \citet[][]{hopkins10b} and \citet[][]{Rodriguez-Gomez2015}. In the next section we will explore how our own model assumptions and treatments for orphans/satellites can impact the merger rate.

\subsubsection{Impact of model assumptions}
\label{sec:modcomp}
\begin{figure*}
	\includegraphics[width=\textwidth]{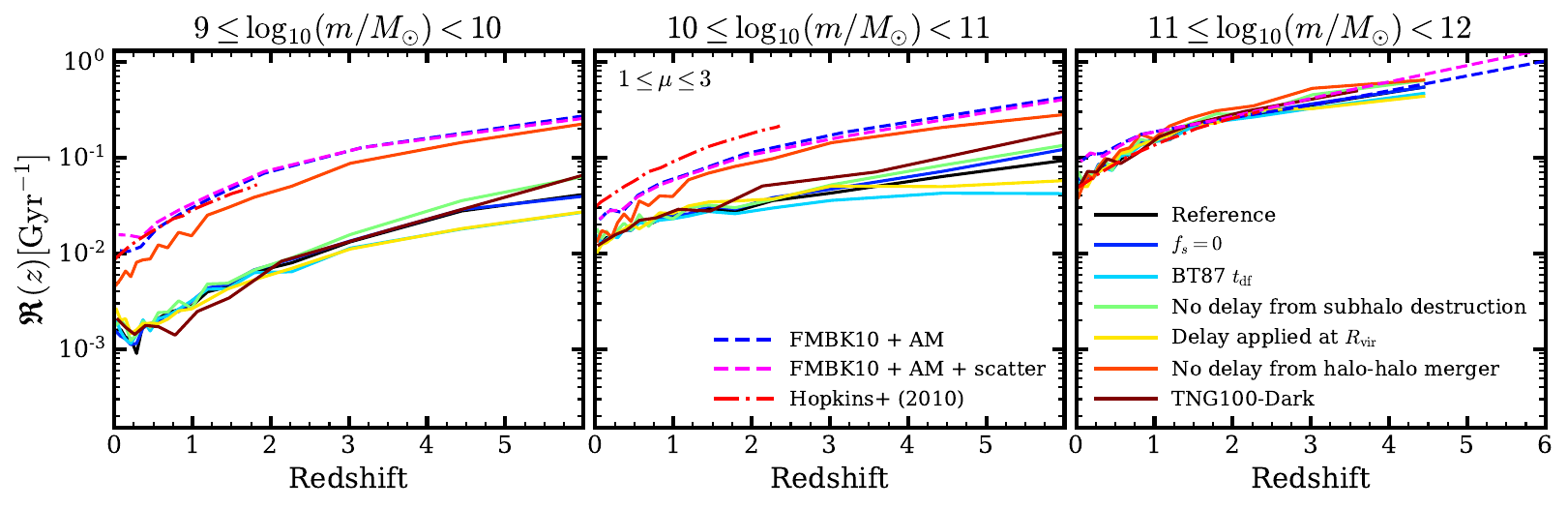}
	\caption{A comparison of major merger rates ($1 \leq \mu \leq 3$) under various model assumptions. Dashed lines indicate merger rates derived using average halo-halo merger rates convolved with a SMF. Solid lines illustrate results derived directly from \textsc{emerge} galaxy merger trees. The solid black line represents the reference results using our standard model shown throughout this paper. Coloured solid lines show our results under different implementations for dynamical friction, satellite evolution, and simulation resolution. Methods using average halo-halo merger rates overpredict galaxy merger rates compared to full tree based methods in all cases. This over-prediction becomes more extreme at lower stellar masses.}
	\label{fig:modcomp}
\end{figure*}

Although the purpose of this work is not to address the systemics that impact the predicted merger rates across the range of available models, we can address known sources of uncertainty, and determine what role our model assumptions play in the resulting merger rates. In this section we will vary core components of \textsc{emerge} and analyse the resulting merger rates to determine if any of our model assumptions are suppressing merger rates compared to other empirical approaches \citep[i.e.][]{Stewart_2009, Hopkins2010}. Throughout this section we take our standard model implementation with the parameters of Table~\ref{tab:best_fit} as our reference. We only explore the impact these model assumptions have on the \textit{major} merger rate with $1\leq\mu\leq3$. We chose this definition of major merger for ease of comparison with \citet{Hopkins2010}. Figure~\ref{fig:modcomp} illustrates the results of this study alongside \citet{Hopkins2010}.  As we vary each model element we do not refit for each model permutation. Consequently, these model variations do not reproduce all observations as accurately as the reference case. Our model permutations cover the following:
\begin{enumerate}
    \item Impact from our satellite stripping implementation. 
    \item Swapping the dynamical friction formula from \citet{BK08} to \citet{BT87} .
    \item No orphan galaxies. Satellites galaxies are merged onto their host when their sub-haloes are lost in the simulation.
    \item All satellites as orphans. Orphan galaxies are initiated when one halo enters the virial radius of another for the first time.
    \item No satellites. Galaxies are merged at the same time as their host haloes.
    \item Increased resolution. Standard \textsc{Emerge} options applied to a simulation with a much higher particle/halo resolution.
    \item Average halo-halo merger rates with rank ordered abundance matching, using the SHMR derived directly from our reference model.
    \item Average halo-halo merger rates and abundance matching with scatter allowed in the SHMR.
\end{enumerate}

In section~\ref{sec:stripping} we described our methodology for stripping satellite galaxies, and our newly implemented approximation for halo mass-loss in orphan galaxies, which presents the possibility that our implementation may strip orphan satellites too strongly, suppressing the merger rate. We can provide a simple check for this scenario by setting the stripping parameter $f_s = 0$. This ensures that all satellites with a short enough dynamical friction will merge and contribute to the computed merger rate. In Figure~\ref{fig:modcomp} this case is illustrated by the blue line labelled `$f_s=0$'. We find a $\sim33$ per cent boost in the major merger rate at intermediate masses, but at low and high masses we see no discernible change in the merger rate. Though not displayed in Figure~\ref{fig:modcomp} we also verified that our merger rates are nearly unchanged within the $f_s\pm1\sigma$ parameter range noted in Table~\ref{tab:best_fit}. Also, setting $f_s=0.1$ did not reduce the merger by more than a factor $\sim2$ compared to the reference, though we note that both the $f_s=0$ and $f_s=0.1$ cases do not reproduce the local clustering data well. We can conclude from these tests that our major merger rates are not strongly impacted by our current implementation for halo mass loss or stripping.

The models shown in Figure~\ref{fig:models} adopt a wide range of dynamical friction formulations. The use of \citet{BK08} dynamical friction has by now become the standard for controlling satellite decay, and builds upon earlier work by tuning the merging timescales using a suite of idealised $N$-body mergers that track haloes and their baryonic components from infall to final coalescence. While this formulation should provide a more physical description for the satellite decay process, we can nonetheless explore merger rate with the more classical dynamical friction formula provided by \citet{BT87}. We find that with respect to major mergers the choice of dynamical friction makes little difference for massive galaxies. At intermediate and low mass we find lower merger rates compared to our reference case. This effect is most pronounced at high redshift and intermediate masses where we see a $\sim56$ per cent difference from the reference.

The need for orphans in modelling has been, by this point, thoroughly discussed \citep[e.g.][]{Campbell2018, Behroozi2019}. Where models continue to differ is in the precise treatment of orphan galaxies, with a core difference being \textit{when} orphans are placed into the simulation. It has been argued that models relying on $N$-body trees that track substructure experience lower merger rates due to overly effective tidal stripping of subhaloes with the absence of baryons \citep{hopkins10b}. Furthermore, it has been argued that applying a dynamical friction recipe at the time of subhalo disruption can introduce additional uncertainty as these formulations are calibrated from the initial halo-halo merger. Some tests have shown that without sufficiently resolved subhaloes this can artificially increase merger timescales by as much as a factor $~8$ \citep{hopkins10b}. Now, we probe whether our treatment for orphans is suppressing the merger rate compared to other methods.

In Figure~\ref{fig:modcomp} the green line labelled `No delay from subhalo destruction' indicates the scenario where we `turn off' orphans and simply merge satellite galaxies with their host when the subhalo is lost in the simulation. This test can best be compared with methods that determine the merger rate by convolving a SHMR with the average subhalo destruction rate \citep[e.g.][]{Stewart_2009, Rodriguez-Puebla2017}. It has been shown that the majority of haloes are disrupted in the inner halo \citep{Wetzel2010}, where the remaining time before final merger is relatively short. Thus by merging galaxies at the time of subhalo disruption we are placing an upper limit on the merger rate compared to our reference case. Indeed we do see that this scenario produces a increase in major merger rates for each mass bin. Overall this contributes as much as a $\sim58$ per cent increase compared to the reference model. From this we can see that by implementing dynamical friction at subhalo loss we are not creating a substantially longer lived population of orphan satellites.

\begin{figure}
	\includegraphics[width=\columnwidth]{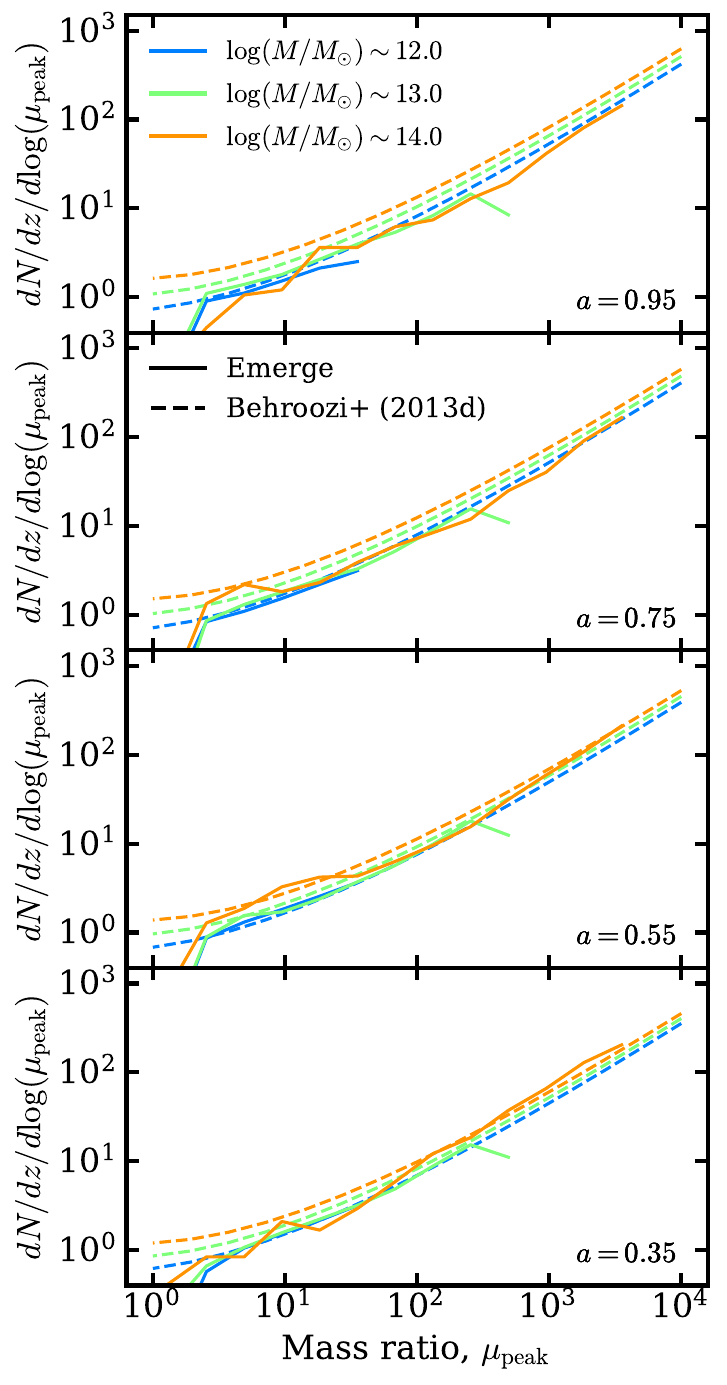}
	\caption{The subhalo destruction rate from our input halo merger trees (solid lines) compared to the best fit formula presented in \citet{Behroozi2013d} (dashed lines). Each panel adopts a uniform binning in scale factor were $\dd a=0.1$. The mass ratio is taken with respect to the host halo mass and the subhalo \textit{peak} halo mass, $\mu_{\mathrm{peak}} \equiv M_{\mathrm{host}}/M_{\mathrm{sub,peak}}$}
	\label{fig:sh_destruction}
\end{figure}

In section Section~\ref{sec:halohalo} we showed that the halo-halo merger rates from our halo merger trees are consistent with the fits shown in \citet{Genel09, FMB10}. As an additional check on this model variation we compare the subhalo destruction rate measured from our input halo merger trees against the fitted relation provided in \citet{Behroozi2013d}. Subhalo disruption is defined when a subhalo can no longer be tracked in the simulation. The mass ratio is taken between the host halo mass and the subhalo \textit{peak} mass. Having shown that removing orphans does not strongly impact our merger rates, one could see how altering the subhalo destruction rate would influence the resulting galaxy merger rate. Figure~\ref{fig:sh_destruction} illustrates the subhalo destruction rate.  Generally, we find that our subhalo destruction rates show a similar mass ratio scaling as \citet{Behroozi2013d}. We can also note that our rates show a weaker scaling with primary halo mass, and tend to be lower when compared to \citet{Behroozi2013d}. Such a result is consistent with the lower galaxy merger rates we exhibit compared to some other models relying on similar methods. However, we have not explored in detail the source of this discrepancy. It is not the goal of this work to probe how strongly this metric is affected by cosmology, definition for peak mass, box size or resolution.

Next, we alter the model such that orphan galaxies are initiated at the virial radius at the time of halo-halo merger. By doing so we can evaluate whether our implementation for dynamical friction is inducing prolonged merging timescales compared to the use case for which the dynamical friction formula was calibrated. To perform this test we remove all subhaloes from the input merger trees. The resulting merger rates from this model are indicated by the yellow line labelled `Delay applied at $R_{\mathrm{vir}}$' in Fiqure~\ref{fig:modcomp}. In this case we actually observe a slightly lower merger rate at all masses compared to the reference. We find a $\sim40$ per cent lower merger rate at intermediate masses by $z=6$.

As a final check on our orphan treatment we run our standard model with a more highly resolved $N$-body simulation. For this test we utilise TNG100-Dark, the dark-matter only run of the Illustris TNG100 simulation \citep{Springel2018}, with merger trees constructed using \textsc{Rockstar} and \textsc{ConsistentTrees}. This simulation uses a Planck cosmology in a periodic volume with side lengths of $110.7$ Mpc. This simulation contains $1820^3$ dark matter particles with particle mass $8.86 \times 10^6 \mathrm{M}_{\odot}$. With this test we can determine whether our major merger rates are significantly affected by resolution. The results from this model are labelled as `TNG100-Dark' in Figure~\ref{fig:modcomp}. We find that our results are robust to a substantially increased mass resolution. While this volume does produce a marginal increase in merger rates at high redshift, we find that the broader trends are fundamentally equivalent, and any difference is within the range of sample variance between the dark matter simulations.

So far the model variations we have introduced have not resulted in extraordinary changes to our standard implementation. To place an upper limit on the merger rates we can expect from this model we explore an extreme case where galaxies are merged at the same time as their parent haloes merge. For this test we once again use our modified halo merger trees where substructure has been removed. This modification results in extreme changes for some model predictions that completely disagree with observational constraints. In particular the SMFs under-predict the abundances of galaxies with $\log_{10}(m/M_{\odot})\lesssim 11.3$, and small-scale clustering all but vanishes. As we are probing the merger rate \textit{per galaxy} the lower abundances can serve as an additional boost to the merger in the affected mass ranges. The end result is that we get higher merger rates at all masses, with the increase in rates becoming most pronounced for low mass galaxies. At the most extreme this approach results in a factor $\sim5$ higher merger rate than the reference case.

Finally, we determine merger rates using the method described in \citet[]{Hopkins2010} but with updated halo-halo merger rates and SHMR to better match \textsc{emerge}. In this approach we derive galaxy-galaxy merger rates by convolving average halo-halo merger rates with the SHMR produced by our reference model SMF via abundance matching. \citet[]{Hopkins2010} showed that this method is robust to changes in dynamical friction, inclusion of substructure, and choice of mass function. In the last decade the quality of observational estimates of galaxy properties has improved substantially; in particular we now have reliable measurements of galaxy properties beyond $z\approx2$, a noted limitation of the abundance matching model they employed for their core model \citep{Conroy2009}. In this final test we investigate if the galaxy merger rate can be constrained using only the average halo-halo merger rate, and observed galaxy mass functions, or if we require the self consistent growth history contained in complete merger trees. 

Beyond the updated SMF we also make a few other changes to the model described in \citet[]{Hopkins2010}. For average halo-halo merger rates we adopt the model presented in \citet{FMB10} along with their best fit parameters. This formulation opts for a more simplified redshift evolution compared to \citet{Fakhouri2008}. The basic procedure for this abundance matching method can be summarised as follows:

\begin{enumerate}
    \item Create mock halo catalogues by sampling the HMF at any redshift for which we want to know the merger rate. We sample our \textit{total} $N$-body HMF directly, enforcing the same halo mass resolution limit as our reference model.
    \item Determine the number of mergers $N(M, z ,\mu_{H})$ for each mock halo. We only consider halo mergers with $\mu_{H}\leq 100$, with redshift bins centred on each `snapshot'.
    \item Compute the dynamical friction time for each halo-halo merger according to \citet{BK08}.
    \item Evolve the main halo mass forward to the time of $t_{df}$. As described in \citet[]{Conroy2009} we use the halo mass growth formula described in \citet[]{Wechsler2002} \
    \begin{equation}
        M_{\mathrm{vir}}(a) = M_{0}\exp\left[-2a_{c}\left(\frac{a_0}{a}-1\right)\right] \; ,
    \end{equation}
    where the average formation scale factor $a_c$is parameterised as
    \begin{equation}
        a_c(M_{\mathrm{vir}}) = \frac{4.1}{c(M_{\mathrm{vir}})(1+z)} \; .
    \end{equation}
    Here $c(M_{\mathrm{vir}})$ is the halo mass-concentration relation at $z=0$. We use the updated form to \citet{Bullock2001}, as presented by \citet{Maccio2008} which takes the form
    \begin{equation}
        c(M_{\mathrm{vir}},z) = K\left[ \frac{\Delta_{\mathrm{vir}}(z_c)}{\Delta_{\mathrm{vir}}(z)}  \frac{\rho_{\mathrm{u}}(z_c)}{\rho_{\mathrm{u}}(z)} \right]^{1/3} \; ,
    \end{equation}
    where $\Delta_{\mathrm{vir}}$ is the overdensity of the halo relative to the mean density of the Universe $\rho_\mathrm{u}$. The parameter $K=3.8$ is the halo concentration at the collapse redshift $z_c$ and is fit to numerical simulations. We allow for $\Delta\log_{10}(c_{\mathrm{vir}})=0.14$ when computing halo growth.
    \item We populate haloes with galaxies at $t_{df}$ using a simple rank ordered abundance matching, using the evolved halo mass for the central galaxy and the infall halo mass for the satellite galaxy.
    \item Finally the merger rate per galaxy is computed using the same strategy described in \ref{sec:intrinsic}.
\end{enumerate}

The results of this test are displayed with dashed lines labelled `FMBK10 + AM', we performed this analysis both with (magenta) and without (blue) scatter in the SMHR. In either case we find that this method over-predicts intrinsic major merger rates at all masses. The difference is most extreme at low masses where there is an order of magnitude over-prediction in the major merger rate. Given this discrepancy we conclude that average halo merger rates are ill-suited for deriving galaxy merger rates.

These differences in predicted merger rate among various models makes make clear that the methodology chosen to link halo and galaxy properties has tangible impact on the assembly pathway of galaxies. In the next section we will utilise mock observations of our simulated galaxy catalogues to gain a better understanding of galaxy assembly in our framework.

\section{Observed merger rates}
\label{sec:obs}
So far we have established the intrinsic galaxy assembly process within the context of our model. The next step is to take this knowledge of galaxy assembly and translate that into something we might observe. Observationally, the galaxy-galaxy merger rate is difficult to ascertain. Additionally, the dynamic process of merging takes place on the scale of hundreds of Myrs to Gyrs.

Obvious physical tracers of a recent merger such as disturbed morphologies present one option for deducing the galaxy merger rate. Methods invoking quantitative morphology such as $G-M_{20}$ or asymmetry are not equally sensitive to all merger mass ratios. Furthermore, these morphological methods are sensitive to total mass, gas properties, orbital parameters, merger stage, and viewing angles \citep[]{Abraham2003, Conselice2003, Lotz2008, lotz2011, Scarlata2007}.  These additional difficulties present a greater barrier to identifying mergers and determining a cosmological merger rate \citep[]{Kampczyk2007, Scarlata2007, Lopez-Sanjuan2009, Shi2009, Kartaltepe2010, Abruzzo2018, Nevin2019}. One common observational method for deriving the galaxy merger rate is through the analysis of galaxies in close pairs. The foundation of this approach is simple, as galaxies found in close proximity are expected to merge within some finite predictable time scale.

Within theoretical models we have the possibility of investigating the complete growth history of galaxies in a cosmological volume, and by performing mock observations on our simulated catalogue we are able to provide guidance on how physical observables can be translated into a true merger rate.  The standard galaxy lists produced with \textsc{emerge} provide an ideal sandbox for comparing observed merger rates with theoretical predictions.

In this section we will be studying two particular quantities: the evolution of the galaxy pair fraction, and the observation timescale of close pairs. The close pair fraction is a measurement of how galaxies cluster, not unlike the projected correlation function $w_p$. While these observables are related they are not directly interchangeable. In particular as we will soon see the pair fraction is used as a proxy for the merger rate and is subject to additional selection criteria to maximise the likelihood that observed pairs are physically associated and expected to merge. Furthermore, \textsc{emerge} is only fit to the stellar mass projected correlation function at $z=0$, wheras the pair fraction must be measured to high redshift. In section~\ref{sec:stripping} we noted model improvements that needed to be made in order to fit small scale clustering. This highlights the fact that this particular observable \textit{is} sensitive to the model. Additionally, in section~\ref{sec:modcomp} we explored several variations of our model and how those changes impact the merger rate. In most cases the resulting SMF under these variations was different from the reference, but largely within observed errorbars. With regards to clustering this is not the case, we find while these changes have only a small influence on galaxy abundances and merger rates, they show far greater influence on galaxy clustering. As we do not refit the model for each of these variations we cannot say whether all models produce the same pair fraction evolution provided an equally good fit to the $z=0$ projected correlation function.

An advantage of empirical modelling is in the `observables first' approach. Because these models match observation by design, they are the ideal test-bed for relating the co-evolution of large scale scale observables with other extrinsic galaxy properties such as their \textit{ex-situ} stellar mass growth. Numerical simulations are a necessary tool in bridging this gap, so in this next section we will see how our model compares with current observations, and whether our results are in agreement with the expectations set by other theoretical models.

\subsection{Close galaxy pairs}
\label{sec:pairs}
\begin{figure*}
	\includegraphics[width=\textwidth]{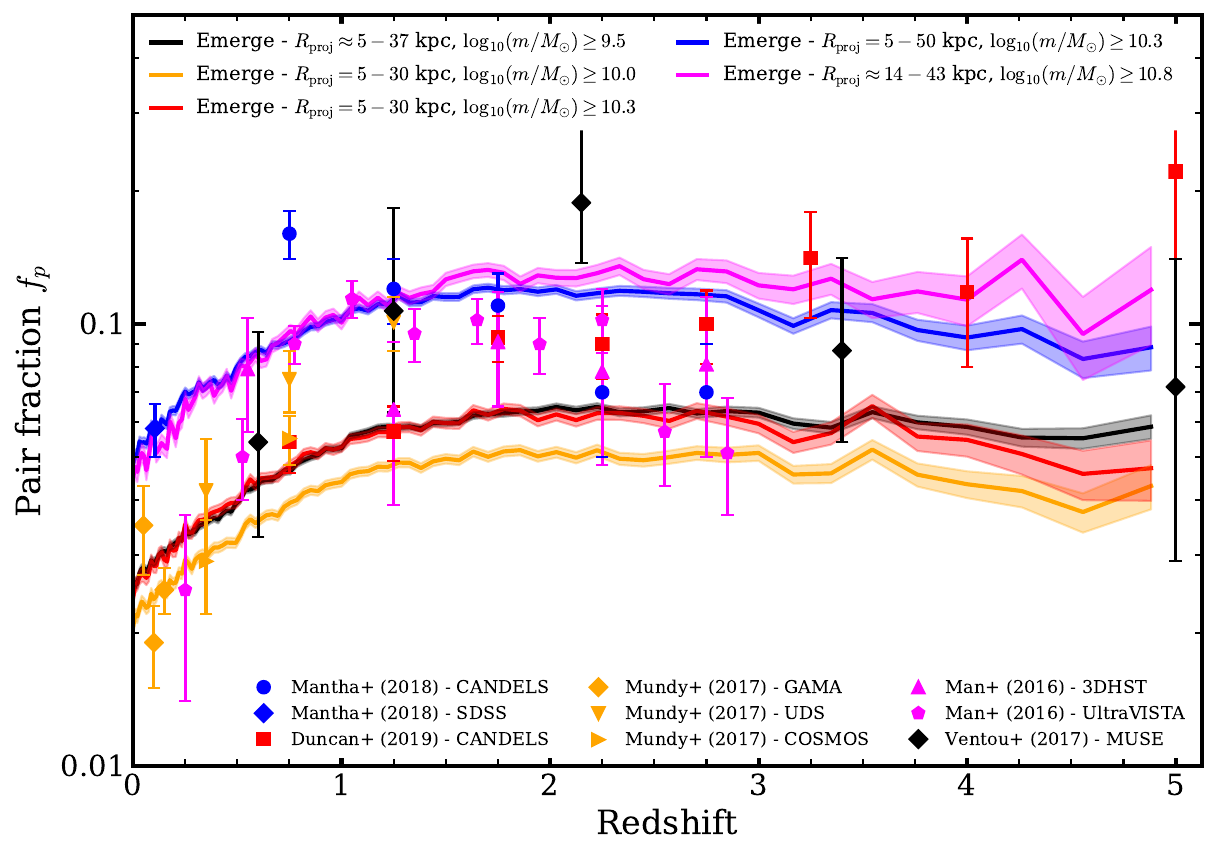}
	\caption{Redshift evolution of the pair fraction. We present the major pair fraction $(\mu \leq 4)$ for our simulation alongside observed pair fractions \citep{Man_2016, Mundy2017, Ventou2017, mantha2018, Duncan2019}. Solid lines illustrate the pair fraction from our model  under four different combinations of radial projected separation ($R_{\mathrm{proj}}$) and mass threshold. The shaded regions indicate Poisson error in the number of pairs. These combinations where chosen to provide the best comparison with observations displayed. Line and marker colours indicate which results share the same selection criteria. Our methods best match observations that incorporate spectroscopic redshift data, these data points are noted by diamond markers. The black line is employs the pair selection criteria of \citet{Ventou2017} with $R_{\mathrm{proj}} = 3-25$ kpc $\mathrm{h}^{-1} \approx 5-37$ kpc, only this work provides spectroscopic data out to $z=5$. The majority of results indicate an increasing pair fraction at low redshifts, and either a flat or decreasing pair fraction at high redshift.}
	\label{fig:pair_fraction}
\end{figure*}

A typical pair count requires two quantities: a projected galaxy separation radius $R_{\mathrm{proj}}$, and some additional redshift proximity criterion. While it is in principle possible to use a 3D deprojected radius to determine physical proximity, this method is prone to error for galaxies with large uncertainties in redshift. When reliable spectroscopic redshift data are available and relative proper motions of companion galaxies can be determined, a common criteria is to use a maximum line of sight velocity difference $\Delta v$ to establish physically associated pairs. Pairs with a small enough relative velocity are assumed to be gravitationally bound and will eventually merge. We use the complete information available in our catalogues to perform such an observation. To maintain the most transferable results we selected pairs by stellar masses, $R_{\mathrm{proj}}$ and $\Delta v$ according to values commonly used by observers \citep[]{lotz2011, Man_2016, mantha2018, Duncan2019}. At each simulation snapshot we compute the fraction of galaxies hosting a major ($\mu\leq4$) companion, $f_p = N_{\mathrm{pairs}}/N_{\mathrm{gal}}$. We do not construct light-cone catalogs, as such our analysis does not incorporate field variance, nor do we impose volume restrictions at low redshift to approximate sample incompleteness from a narrow field. Our analysis only considers major galaxy pairs, where the most massive galaxy in each pair must reside above a specified mass threshold (Table~\ref{tab:fp_fit} indicates the mass thresholds used in this work). All measurements adopt a fixed $\Delta v = 500\, \mathrm{km} \, \mathrm{s}^{-1}$ for redshift proximity, consistent with previous works \citep[]{Patton2000, Lin2004, Lin2008, deRavel2009, lotz2011, Lopez-Sanjuan2012, mantha2018, Mundy2017}.

In Figure \ref{fig:pair_fraction} we display the pair fraction of our simulation alongside recent observations. In this figure we show our pair fractions using five different criteria for mass threshold and projected separation. In all cases, regardless of the radial separation chosen, our results express similar features. In each case we see an increase in the pair fraction with redshift, with a peak at $z\approx2.5$, followed by a shallow decrease in pair fraction toward even higher redshifts. Previously published observed pair fractions have shown a power law increase with redshift \citep[]{Kartaltepe2007, Lin2008, Bundy2009, Conselice2009, deRavel2009, Lopez-Sanjuan2009, Lopez-Sanjuan2013, Shi2009, Xu2012}, while more recent observations indicate a flattening or even decreasing pair fraction at higher redshifts \citep[]{Man_2016, Mundy2017, Ventou2017,Ventou2019, mantha2018}. In this respect our results more closely align with more recent works. However, the precise functional form remains a point of contention. We find that our pair fractions are most appropriately fit with a modified power-law exponential function \citep[]{Carlberg1990, Conselice2008}:
\begin{equation}
	f_{p}(z) = \alpha(1+z)^m\exp[\beta(1+z)] \;.
	\label{eq:fp}
\end{equation}

When comparing our results to observations or other models it is important to note some of the inconsistencies that might prevent a more accurate comparison. We chose our pair selection criteria to provide the most direct comparison possible with observations. Though we find qualitatively similar results between our selected apertures, the differences produced are immediately noticeable. The pair fraction is sensitive to the selection criteria applied and in the case of observations, sensitive to the methods used to account for sample completeness. Additionally, in this work we only compare with fractions derived using stellar mass and stellar mass ratio of pairs. Previous works have shown that pair fractions determined using flux ratio, or baryon mass ratio pairs, produce results very different results than stellar mass selected pairs \citep{lotz2011, Man_2016}. Furthermore, observations often lack robust spectroscopic redshift data, instead relying on photometric redshifts. Under the best circumstances scatter in photometric redshift estimates is $\delta z/(1+z)\approx0.01$ \citep[]{Molino2014, Duncan2019}. This level of precision is insufficient to determine relative velocity differences down to $\Delta v = 500\, \mathrm{km} \, \mathrm{s}^{-1}$. Instead of using relative velocities between galaxies, photometric redshift differences along with their associated uncertainties are utilised. One approach is to use $\Delta z^2 \leq \sigma_{1}^2 + \sigma_{2}^2$, where the $\sigma_{1}$ and $\sigma_{2}$ are the photometric redshift uncertainties for the major and minor galaxy in each pair, respectively \citep[]{Bundy2009, mantha2018}. Otherwise, probabilistic methods can also be employed to determine physically associated pairs \citep[]{Lopez-Sanjuan2015, Mundy2017, Duncan2019}. We can see the impact of many of these differences in pair counting methodology if we compare the pair fractions derived from the same field data (\citealt{mantha2018}, blue circles; \citealt{Duncan2019}, red squares). While both of these analyses are based on the same underlying image data, they come to very different conclusions regarding both the normalisation and functional form of the pair fraction.

Considering these difficulties in measuring the pair fraction, the most comparable set of observations for our results are those from \citet{Ventou2017} based on MUSE \citep{Bacon2010} data, who have spectroscopic data out to high redshift. To make a direct comparison with their work we adopt their pair selection criteria where $R_{\mathrm{proj}} = 3-25$ kpc $\mathrm{h}^{-1}$, and $\log_{10}(m/M_{\odot})\geq9.5$ (black line). Although we are in reasonable agreement with \citet{Ventou2017} we note that the MUSE fields are very small. Consequently their uncertainty due to cosmic variance is large, ranging from $\sigma_{v}=0.15$ at $z\approx 0.6$ up to $\sigma_{v}=0.52$ by $z\approx 5$ \citep[]{Moster2011}. The small field size also results in a limited pair sample; in the redshift range where our results disagree the most $1.5\lesssim z \lesssim 3$, only 9 pairs were observed in a sample of 152 galaxies. These large uncertainties in the observed data hinder a more detailed study to describe the differences with respect to our analysis. Similarly, we are able to make a direct comparison to the low redshift SDSS \citep{SDSS} data point presented in \citet{mantha2018}, as well as the low redshift GAMA data point presented in \citet{Mundy2017}. In these instances we are once again in close agreement where spectroscopic redshifts are available.

In Figure~\ref{fig:pair_fraction} we displayed select results most comparable to some recent observational and theoretical predictions. A short summary of the of the pair selection criteria for the data of Figure~\ref{fig:pair_fraction} can be found in Table~\ref{tab:fp_obs}. Table~\ref{tab:fp_fit} provides the best-fit to our simulated pair fractions for an additional set of stellar mass thresholds and $R_{\mathrm{proj}}$. In each of these fits we assume the functional form of eq.~\ref{eq:fp}.

\subsection{The merger rate from close pairs}
\begin{figure*}
	\includegraphics[width=\textwidth]{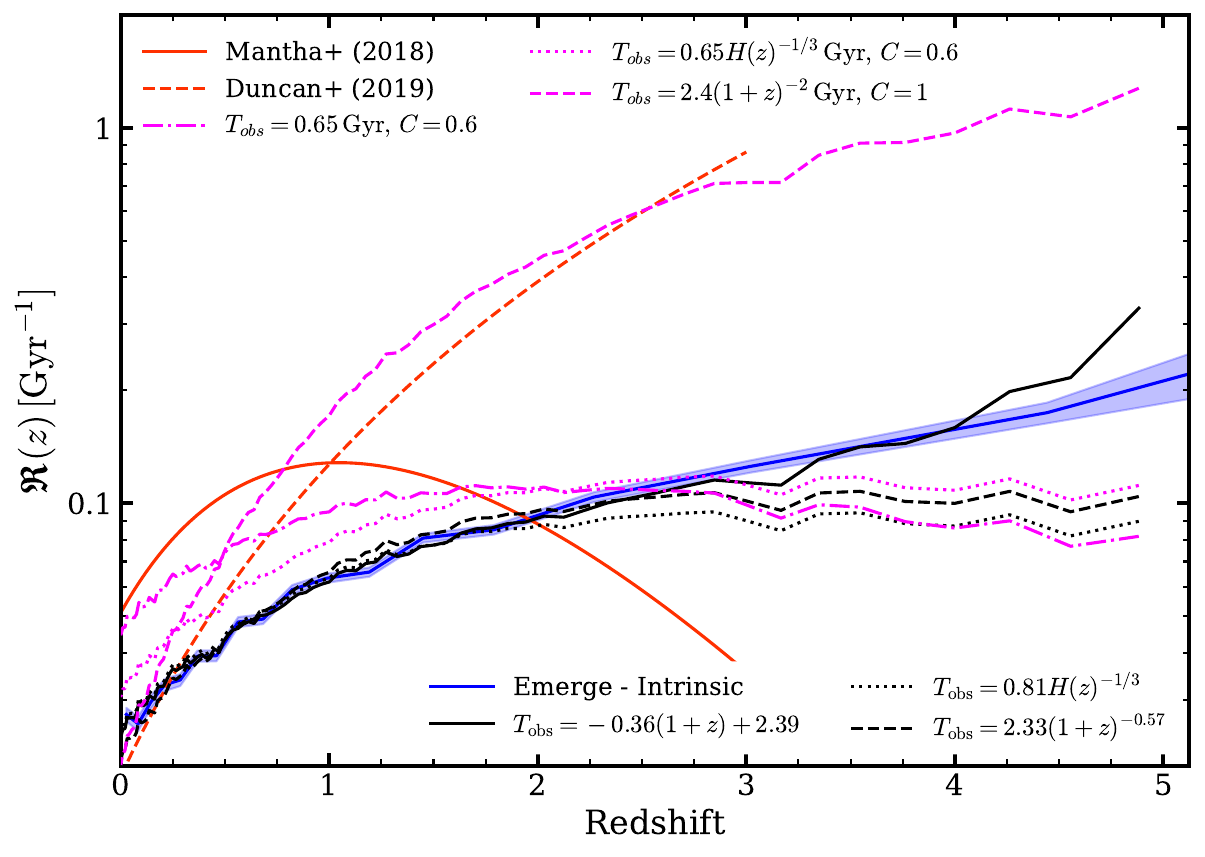}
	\caption{A comparison of observed merger rates (red lines) alongside the merger rate produced through the mock observations shown in Fig.~\ref{fig:pair_fraction}. Our results assume $\log_{10}(m/M_{\odot}) \geq 10.3$, $1 \leq \mu \leq 4 $ and $R_{\mathrm{proj}} = 5-50$ kpc. Magenta lines show merger rates derived from our pair fractions assuming previously published results for $T_{\mathrm{obs}}$. In particular $T_{\mathrm{obs}}=0.65$ \citet{lotz2011}, $T_{\mathrm{obs}}\propto H(z)^{-1/3}$ \citet{Jiang2014}, and $T_{\mathrm{obs}}\propto (1+z)^{-2}$ \citet{Snyder2017}. Black lines illustrate our predicted results from pair fractions under different best fit forms for $T_{\mathrm{obs}}$. The solid blue curve is the underlying intrinsic merger rate produced using the methods described in Section \ref{sec:galaxy-galaxy}, and the shaded blue region depicts poisson error in the merger count. Deriving merger rates using our pair fractions and published values for $T_{\mathrm{obs}}$ results in a poor reproduction of our intrinsic values. Fitting with $T_{\mathrm{obs}}\propto H(z)^{-1/3}$ or $T_{\mathrm{obs}}\propto (1+z)^{m}$ provide a better reproduction of merger rates, but under-predict the intrinsic vales for $z\gtrsim2.5$. Our model intrinsic merger rate is most accurately reproduced assuming a $T_{\mathrm{obs}}$ with a linear redshift scaling, solid black line.}
	\label{fig:pair_rates}
\end{figure*}

Determining a merger rate from pair fractions is \textit{conceptually} straight forward. Observed pairs are assumed to result in a merger on some finite time scale. Therefore, one (mathematically) simple approach to convert a measured pair fraction to a rate is to simply divide the pair fraction by an average observation timescale $T_{\mathrm{obs}}$. Which specifies the amount of time that a pair \textit{could} be identified by the established pair selection criteria. The actual timescale needed for a given galaxy pair to merge can depend on properties other than stellar mass, and projected radial separation. To account for the possibility that not all pairs will merge in the expected timescale or that some pairs are a result of chance projection, an additional correction factor $C_{\mathrm{merge}}$ is often introduced to specify which fraction of the observed pairs will actually end up merging. These quantities can be combined, resulting in the merger rate per galaxy formulation:
\begin{equation}
	\mathfrak{R} = \frac{C_{\mathrm{merge}} \times f_{\mathrm{p}}}{\langle T_{\mathrm{obs}} \rangle} [\mathrm{Gyr}^{-1}]\;.
\end{equation}
This formulation is contingent upon having a pair selection criteria that does adequately identify physically associated galaxy pairs in the early stages of a merger. Additionally, it assumes that adopting an average observation timescale is a suitable method for converting a sample of galaxy pairs into a rate. Under this formulation the observation timescale is a crucial quantity in translating pair fractions to merger rates. Work seeking to characterise this quantity remains in tension regarding the functional form. A common approach is to take $T_{\mathrm{obs}}$ as a constant. Suggested values for a range of stellar masses and $R_{\mathrm{proj}}$ have been proposed by \citet{lotz2011}. Conversely, recent work has suggested formulations for a redshift dependent observation timescale. For instance, \citet{Snyder2017} have proposed $T_{\mathrm{obs}}\propto(1+z)^{-2}$, while \citet{Jiang2014} suggest $T_{\mathrm{obs}} \propto H(z)^{-1/3}$.

In Figure \ref{fig:pair_rates} we compare rates derived from two recent observational results \citep[][red lines]{mantha2018, Duncan2019} to our intrinsic merger rates (i.e. the true merger rate measured in our simulation) and our mock pair fraction derived rates. For clarity we show our results only for $\log_{10}(m/M_{\odot})\geq10.3$, $1\leq\mu<4$, and $R_{\mathrm{proj}}=5-50$ kpc. This aligns with the selection criteria of the displayed observations. The pair selection criteria and $T_\mathrm{obs}$ scaling for the observed data can be found in Table~\ref{tab:fp_obs}. Here, low redshift results agree within a factor of $\sim 2$. However, predictions deviate heavily towards higher redshift. By $z=3$ there is as much as an order of magnitude difference between predicted and observed major merger rates. Generally, these recently published observations over-predict the merger rate compared to our intrinsic values for $z \gtrsim 0.5$. While these methods draw from the same fields, they come to very different conclusions regarding pair fraction evolution. Additionally, different observation timescales are adopted as the default choice to produce each result . If we apply the same observation timescales used in these works (magenta lines) we are unable to reproduce our intrinsic values, from our pair fractions. To better understand these deviations we instead fit $T_{\mathrm{obs}}$ using these proposed formulations. From here we can determine if any can provide a meaningful mapping of the pair fraction into the intrinsic merger rate based on our results. 

Directly comparing the redshift evolution for merger rates and the pair fraction we can see that a constant value for $T_{\mathrm{obs}}$ is insufficient. The increasing merger rate and decreasing pair fraction at high z require that the observation timescale decrease with increasing redshift. If we impose $T_{\mathrm{obs}}\propto H(z)^{-1/3}$ \citep[]{Jiang2014} we find that we are able to reproduce the intrinsic merger rate until $z\sim2$ beyond which the predicted merger rate flattens, under-predicting the intrinsic rate. Similarly, the best fit power-law scaling $T_{\mathrm{obs}}\propto (1+z)^\alpha$ reproduces the low redshift merger rate scaling, but again flattens and under-predicts the merger rate at high $z$. We find that the most simple scaling that can recover the intrinsic merger rate to high $z$ is linear, $T_{\mathrm{obs}} = w(1+z)+b$. Additionally, we find that such a scaling provides a better fit for high stellar mass galaxies than for low stellar masses. In the case of lower masses the linear fit begins to deviate for $z>4$. However, it's clear that this formulation could fail at any mass if the best fit values result in a negative $T_\mathrm{obs}$ at the desired redshift.

Absent a generalised fitting formula, we find that for any mass threshold and mass ratio our intrinsic merger rates and pair fractions can be well fit by a power-law exponential as eq.~\ref{eq:fp}. Table~\ref{tab:fp_fit} shows our best fit intrinsic merger rates, pair fractions, and observations timescales for a few common stellar mass thresholds.

Our findings conflict with those recent works suggesting a strong redshift evolution for observation timescales. In the case of \citet[]{Snyder2017} the proposed scaling where $T_{\mathrm{obs}}\propto(1+z)^{-2}$ provides a mapping from a flat pair fraction to an underlying merger rate that scales as a power-law with increasing redshift. However, as noted in \citet[]{Snyder2017}, the measured pair fractions from their work rely on a mass ratio calculated using galaxy properties at the same redshift for which the mock observation was performed, while the intrinsic merger rate as measured by \citet[]{Rodriguez-Gomez2015} takes the mass ratio with respect to the \textit{peak} stellar mass of the secondary galaxy. This discrepancy in mass ratio definitions makes a direct translation between the intrinsic merger rate and the measured pair fraction troublesome. Subsequently, their proposed scaling for the observation timescale is not necessarily reflecting a physical mechanism driving such a formulation.

When finding the best fit value for $T_{\mathrm{obs}}$ we assume $C_{\mathrm{merge}}=1$. An accurate determination of $C_{\mathrm{merge}}$ is beyond the scope of this work, thus the best fit observation time scale represents an upper limit to the true underlying value. Furthermore, our analysis does not perform a complete light cone analysis, nor do we attempt to reproduce any observational uncertainties in our redshifts or stellar masses. All fits are performed assuming Poisson error in the number of pairs or number of mergers. We leave a more detailed description and analysis of $T_{\mathrm{obs}}$ and $C_{\mathrm{merge}}$ to future works.

\begin{table*}
\centering
\caption{Summary of selection criteria for observed pair fractions. Observations that use a have `CDF' as their redshift proximity indicate a cumulative probability that two galaxies are a pair. Where observations are included in Figure~\ref{fig:pair_rates} we list the redshift depended observation timescale used to translate observed pair fractions into merger rates.}
\label{tab:fp_obs}
    \begin{tabular}{lcccr}
    \hline
     Publication        & $\log_{10}(m/M_{\odot})$              & $R_{\mathrm{proj}}$ [kpc] & Redshift proximity & $T_{\mathrm{obs}}$ [Gyr]\\
     \hline\hline
    Man+ (2016)        & $\geq10.8$                            & $\sim 14-43$              & $\Delta z_{\mathrm{photo}}<0.2(1+z_1)$                            & - \\
     Mundy+ (2017)      & $\geq10.0$                            & $5-30$                    & $CDF(z_1, z_2)$                                                   & - \\
     Ventou+ (2017)     & $\geq9.5 $                            & $\sim 5-37$               & $\Delta v \leq 500$ km/s                                          & - \\
     Mantha+ (2018)     & $\geq10.3$                            & $5-50$                    & $\Delta z_{\mathrm{photo}}^2 \leq \sigma_{1}^2 + \sigma_{2}^2$    & $0.65$ \\
     Duncan+ (2019)     & $\geq10.3$                            & $5-30$                    & $CDF(z_1, z_2)$                                                   & $2.4(1+z)^{-2}$ \\
     \hline
    \end{tabular}
\end{table*}

\begin{table*}
	\centering
	\caption{Best-fit parameters for pair fractions and intrinsic merger rates following the functional form of eq.~\ref{eq:fp}. These best fit values correspond to commonly used mass selections and radial projections used by observers. Rows without a noted $R_{\mathrm{proj}}$ or $T_{\mathrm{obs}}$ are fit to the intrinsic merger rate for the indicated mass. The values assume major mergers only with $\mu\leq4$. listed $T_{\mathrm{obs}} = w(1+z) + b$ are best fit assuming $C_{\mathrm{merge}}=1$.}
	\label{tab:fp_fit}
	\begin{tabular}{lcccccr} 
		\hline
		$\log_{10}(m/M_{\odot})$    & $R_{\mathrm{proj}}$ [kpc] & $\log_{10}(\alpha)$        & $m$             & $\beta$          & $w$    & $b$ \\
		\hline
		\hline
		\multirow{4}{*}{$\geq9.7$}  & $5-30$                    & $-1.489\pm0.006$ & $1.827\pm0.037$ & $-0.542\pm0.017$ & $-0.185\pm0.006$ & $1.528\pm0.024$         \\
		                            & $14-43$                   & $-1.325\pm0.006$ & $1.824\pm0.036$ & $-0.565\pm0.017$ & $-0.262\pm0.008$ & $2.139\pm0.036$        \\
		                            & $5-50$                    & $-1.154\pm0.005$ & $1.800\pm0.031$ & $-0.552\pm0.015$ & $-0.498\pm0.026$ & $3.446\pm0.092$        \\
		                            & $-$                       & $-1.785\pm0.018$ & $1.277\pm0.127$ & $-0.100\pm0.055$ & $-$              & $-$       \\
		\hline
		\multirow{4}{*}{$\geq10.3$} & $5-30$                    & $-1.333\pm0.006$ & $1.972\pm0.034$ & $-0.619\pm0.016$ & $-0.177\pm0.007$ & $1.205\pm0.023$        \\
		                            & $14-43$                   & $-1.194\pm0.007$ & $1.955\pm0.041$ & $-0.643\pm0.020$ & $-0.241\pm0.009$ & $1.586\pm0.030$        \\
		                            & $5-50$                    & $-1.022\pm0.005$ & $1.933\pm0.033$ & $-0.630\pm0.016$ & $-0.360\pm0.013$ & $2.385\pm0.045$        \\
		                            & $-$                       & $-1.530\pm0.020$ & $1.519\pm0.134$ & $-0.160\pm0.059$ & $-$              & $-$        \\
		\hline
		\multirow{4}{*}{$\geq11.0$} & $5-30$                    & $-1.143\pm0.014$ & $2.166\pm0.082$ & $-0.561\pm0.040$ & $-0.126\pm0.007$ & $0.865\pm0.019$        \\
		                            & $14-43$                   & $-1.000\pm0.015$ & $2.270\pm0.084$ & $-0.657\pm0.042$ & $-0.171\pm0.007$ & $1.097\pm0.024$        \\
		                            & $5-50$                    & $-0.832\pm0.009$ & $2.172\pm0.055$ & $-0.607\pm0.027$ & $-0.257\pm0.012$ & $1.666\pm0.037$        \\
		                            & $-$                       & $-1.189\pm0.028$ & $1.774\pm0.186$ & $-0.133\pm0.085$ & $-$              & $-$        \\
		\hline
	\end{tabular}
\end{table*}

\section{Which galaxies grow through mergers?}
\label{sec:history}

In this last section we will move on from addressing the galaxy-galaxy merger rate to exploring the role that mergers play for the growth of galaxies. We aim to answer two key questions:
\begin{itemize}
	\item Where does a galaxy's stellar mass come from?
	\item Are all types of mergers equally important?
\end{itemize}
We approach these questions in the context of the main branch evolution of the $z=0$ galaxy population. Here we explore the merging history of individual galaxies.

\subsection{Stellar mass fraction accreted through different merger types}
\label{sec:exsitu}
\begin{figure}
	\includegraphics[width=\columnwidth]{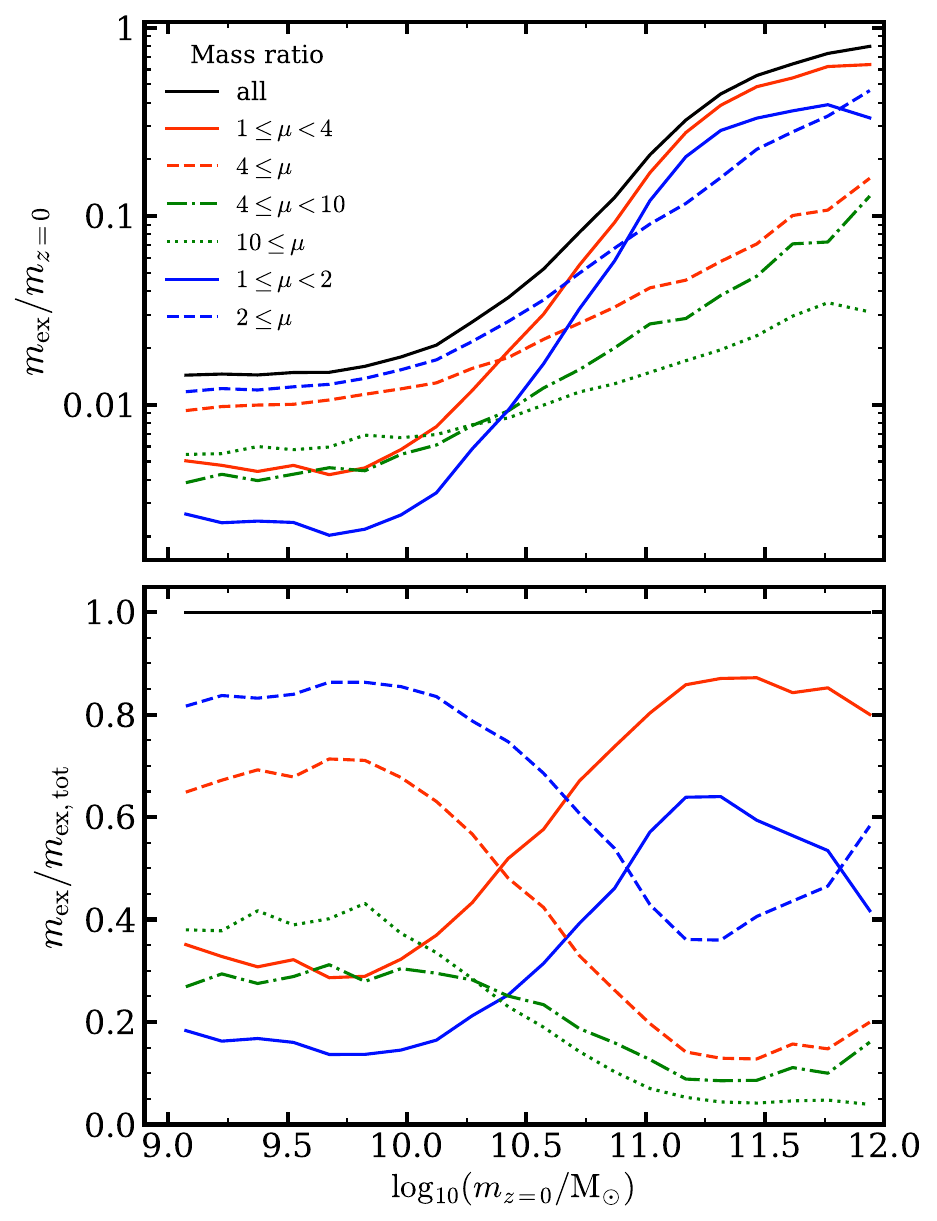}
	\caption{Contribution of accreted mass by merger ratio as a function of $z=0$ stellar mass $m_{z=0}$. \textit{Top panel}: The ex-situ mass fraction $(m_{\mathrm{ex}}/m_{z=0})$ with respect to the $z=0$ stellar mass. The solid black line represents the average accreted mass fraction across all $z=0$ galaxies with $\log_{\mathrm{10}}(m/M_{\odot}) \geq 9.0$ within the simulation volume. Lines of different colour and line style illustrate the average accreted mass fraction broken down by merger mass ratio $\mu$. Lines of the same colour sum to the total accreted mass fraction, solid black line. \textit{Bottom panel}: The fraction of the $z=0$ accreted stellar mass, $m_{\mathrm{ex,tot}}$ broken down by merger mass ratio. Similar to the top panel, line type and colour show the contributions by merger mass ratio with like colours summing to the total average. For instance, the solid blue line illustrates the fraction of all accreted mass deposited through mergers with $1\leq \mu \leq 2$. At the lowest mass this means that only $\sim20$ per cent of accreted material is deposited by mergers of this type. Conversely, the dashed blue line shows the fraction contributed by mergers $\mu \geq 2 $. As these two scenarios represent complete accretion history these lines sum to $m_{\mathrm{ex}}/m_{\mathrm{ex,tot}}=1$, the solid black line.}
	\label{fig:exsitu}
\end{figure}
First we investigate the assembly of galaxies and whether the stellar mass has grown mainly through star formation (in-situ) or through mergers (ex-situ). Previous work has shown that the accreted stellar mass fraction, $f_{\mathrm{acc}}$, ranges between less than $2$ per cent for low-mass galaxies to more than $50$ per cent for massive galaxies \citep[]{Lackner2012, Cooper2013, Lee2013, Rodriguez-Gomez2015a}.

Figure \ref{fig:exsitu} illustrates the fraction of accreted material as a function of $z=0$ stellar mass, divided into different merger mass ratios. The top panel illustrates the accreted mass fraction with respect to total stellar mass, the bottom panels shows the accreted fraction delivered by merger type with respect to the total \textit{accreted} stellar mass. In the previous sections we showed that for the global merger rate more massive galaxies are biased to experience major mergers, while low-mass galaxies are biased to experience mostly mini mergers. Massive galaxies will pass through both of these regimes during their lifetime, so  we can reasonably ask ourselves which mergers ultimately built the galaxies we see? Were galaxies quickly assembled through successive minor mergers, or are the most massive galaxies assembled (late) through major mergers?

Looking at the top panel of Figure~\ref{fig:exsitu} we first find that the largest galaxies are constructed primarily through accreted material. On average we expect upwards of $f_{\mathrm{acc}}\approx80$ per cent at the massive end, and as little as $f_{\mathrm{acc}}\approx1.5$ per cent at the low-mass end for $z=0$ galaxies. There exists a strong mass dependence in the accreted mass fraction between $\log_{10}(m/M_{\odot})=10.25$ and $\log_{10}(m/M_{\odot})=11.25$. In this regime we see a corresponding inversion in the relative contribution of major and minor mergers to the final system. Looking at the solid red line we can see clearly that the most massive galaxies are indeed assembled by successive major merging events. On the massive end, we see that these major mergers contribute as much as $90$ per cent to the total accreted mass budget of a galaxy (bottom panel). Additionally, find that these mergers begin to dominate the accreted mass budget at around $\log_{10}(m/M_{\odot})=10.3$. If we look at the classical merger mass ratio definitions, we see that up until $\log_{10}(m/M_{\odot})=10$, major ($\mu<4$), minor ($4 \leq \mu < 10$), and mini ($10 \leq \mu$) mergers contribute roughly equal quantities to the total accreted mass budget. Beyond this point we see the relative contributions diverge. This contrasts strongly with recent results that indicate major mergers contribute a roughly flat $50$ per cent of the accreted mass fraction at all mass scales, and minor/mini mergers show a roughly constant $20$ per cent contribution \citep{Rodriguez-Gomez2015a}.

Understanding the source of accreted stellar material can have direct consequences on the internal kinematics, and stellar mass distributions of a galaxy. Some models indicate that major mergers deposit stellar material at the center of the descendant galaxy \citep{Deason2013, Pillepich2015}, while minor mergers tend to deposit material at larger radii, growing the stellar halo \citep[]{Hilz2012, Hilz2013, Karademir2019}. Subsequently observations of stellar populations could be used to determine the merging history of galaxies \citep[]{Merritt2016, Amorisco2017, Amorisco2019, Bernardi2019, Ferre-Mateu2019, Hendel2019}.

\begin{figure}
	\includegraphics[width=\columnwidth]{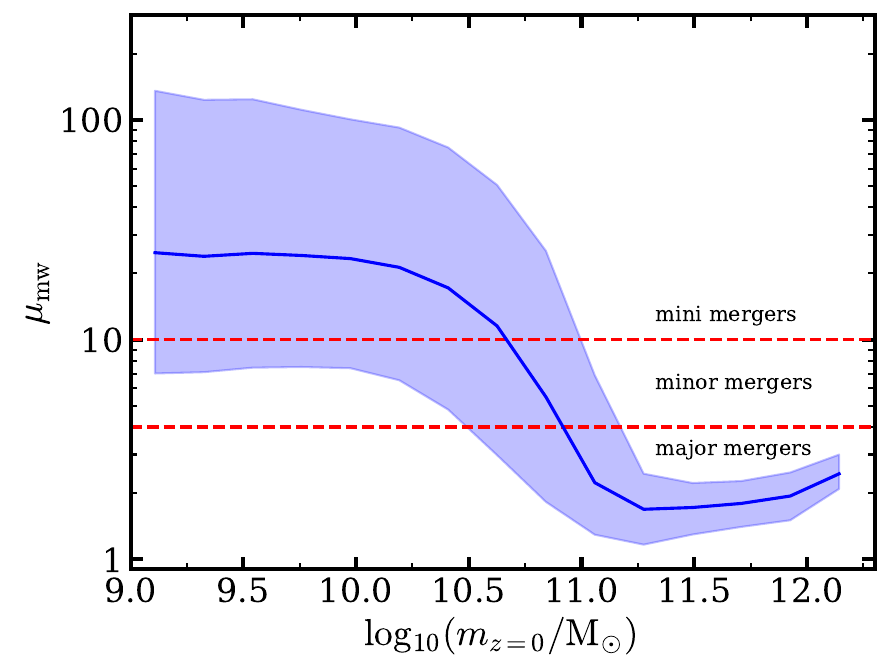}
	\caption{The mass-weighted mass-ratio as a function of $z=0$ stellar mass $m_{z=0}$. The solid line illustrates the median mass weighted mass ratio of mergers along the galaxy main branch for each galaxy in the specified mass range that experienced at least one merger in its lifetime. The shaded region shows the $68^{th}$ percentile surrounding the median.}
	\label{fig:mwmr}
\end{figure}

We can more directly probe which mergers contribute the most to the stellar mass growth of a galaxy by evaluating the \textit{mass-weighted} mass ratio defined as:
\begin{equation}
	\mu_{\mathrm{mw}}\equiv \frac{\sum_{i}^{N_{\mathrm{m}}} m_{i,2}}{\sum_{i}^{N_{\mathrm{m}}} \frac{m_{i,2}}{m_{i,1}} m_{i,2}}
\end{equation}
In this approach, each merger has its mass-ratio weighted by the amount of stellar mass contributed to the final system. This way we see what types of mergers were on average most important for the growth of galaxies at a given mass scale. In Figure~\ref{fig:mwmr}, we show the median $\mu_{\mathrm{mw}}$ for all $z=0$ galaxies in our simulation. In agreement with the results displayed in Figure~\ref{fig:exsitu}, we see that low-mass galaxies on average experience mergers with $\mu_{\mathrm{mw}}\approx 40$, once again illustrating that major mergers are not important for the growth of low-mass galaxies. Previous works \citep[]{Naab2009, Oser2012, Hilz2012, Hilz2013} have suggested that successive minor mergers can be an effective pathway to form large galaxies. Figure~\ref{fig:mwmr} shows clearly that there is only a narrow transition region where these minor mergers play a significant role in stellar mass assembly. For massive systems we once again see that most of the stellar mass is delivered through major mergers. We can see that on average the mergers that bring the most mass into the system are very major mergers with $\mu\approx2$.

\subsection{How frequent are different kinds of mergers?}
\begin{figure*}
	\includegraphics[width=\textwidth]{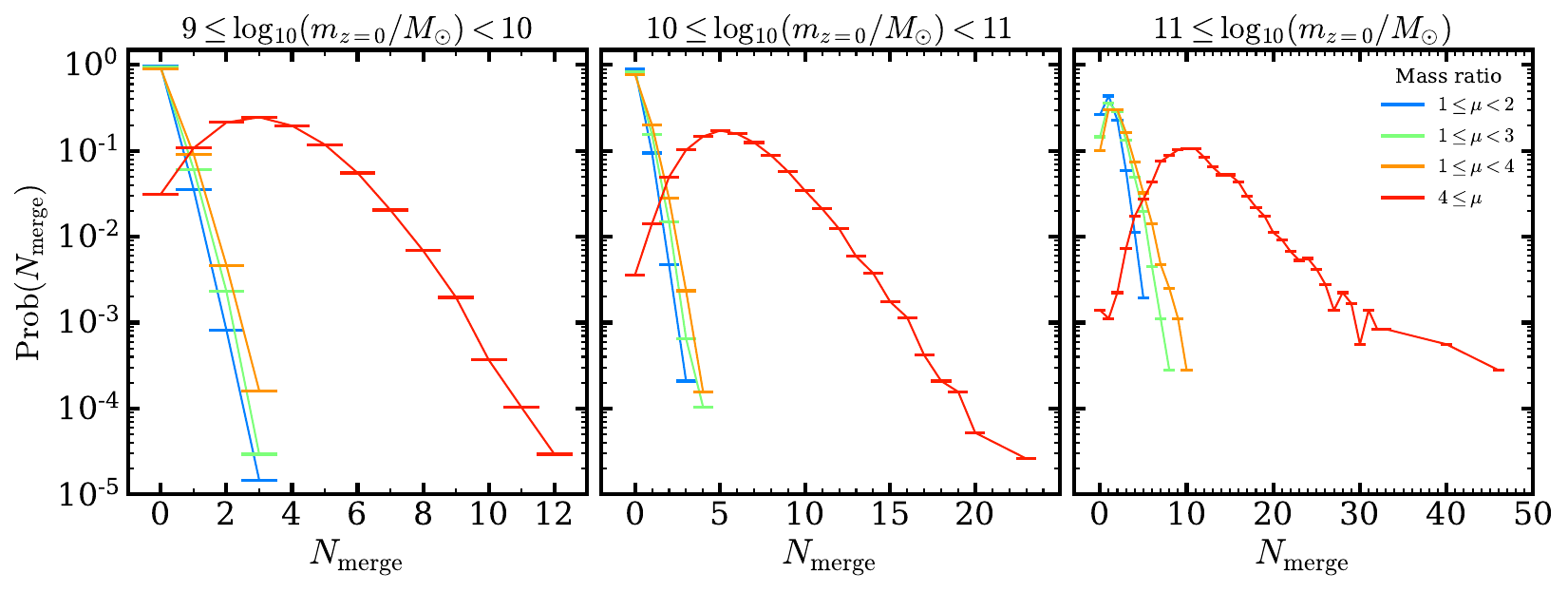}
	\caption{Probability of experiencing $N$ mergers in a $z=0$ galaxy's history. Each line shows the number of mergers occurring along the main branch per $z=0$ galaxy. Each line sums to 1, representing the merging history of all $z=0$ galaxies. For galaxies with $\log_{10}(m_{z=0}/M_{\odot}) < 11$ more than $90$ per cent of galaxies do not experience any major mergers $\mu < 4$. For galaxies with $\log_{10}(m_{z=0}/M_{\odot}) \geq 11$ only $\sim10$ per cent of galaxies do not experience any major mergers.}
	\label{fig:lifetime_mergers}
\end{figure*}

So far this work has focused on the average merger rate, or the average merging history across an entire population of galaxies in a cosmological volume. In this section we investigate the individual merging history of galaxies to see how many mergers a galaxy has experienced in its (main-branch) lifetime.

Figure \ref{fig:lifetime_mergers} shows the number of mergers experienced along the main branch per galaxy. This accounts for the complete main branch merging history of each $z=0$ galaxy in the simulation and indicates the probability distribution of a galaxy having a given number of mergers with a certain mass ratio. Similar to our merger rate analysis we separate the galaxies into three stellar mass bins. In the lowest mass bin (left panel) we see that more than $90$ per cent of galaxies experience no major mergers along their main branch, consistent with the expectations set in the previous section, though more than $85$ per cent of galaxies in this mass bin will experience \textit{at least} one merging event in their lifetime.

At intermediate masses (central panel), we start to observe the larger frequency of major mergers. However, we find that under the loosest definition of a major merger, more than $80$ per cent of galaxies will experience no major mergers in their lifetime. This suggests that a galaxy like the Milky Way has a low likelihood of having ever been impacted by a major merger. Probing in the more narrow range $10.6 \leq \log_{10}(m/M_{\odot}) < 10.8$ we still find that $\sim70$ per cent of galaxies will have no major merger in their lifetime. Conversely, we find that the majority ($97$ per cent) will have experienced \textit{some} merger in their life, no matter how small. This is in agreement with recent observations \citep[]{Dsouza2018, Helmi2018}.

This trend changes for the highest mass bin where mergers play a substantial role in stellar mass growth. In this range we can see that $90$ per cent of galaxies will experience \textit{at least} one major merger in their lifetime, with one galaxy experiencing as many as 11 major mergers. This particular galaxy is the largest galaxy in our box with a stellar mass of $\log_{10}(m_{z=0}/M_{\odot}) = 12.17$, and had already grown to $\log_{10}(m/M_{\odot}) \approx 11$ before encountering its first major merger (near $z\approx 2.5$). However, on average the most massive galaxies experience $~\sim 1.8$ major mergers in their lifetime.

While mergers play very different roles for the evolution of low-mass and massive galaxies, we find that most galaxies are subject to a merger in their lifetime. For $z=0$ galaxies with $\log_{10}(m/M_{\odot})\geq9.0$ we find that $\sim92$ per cent experience a merger with any mass ratio, while $\sim8$ per cent of galaxies experience no merging event along their main branch. However, the complete history of a galaxy is more complicated. Here we have only probed galaxy mergers along the main branch for the most evolved galaxies. This is \textit{not} a complete accounting of the number of mergers that occurred within a galaxy's complete evolutionary tree.

\section{Discussions \& Conclusions}
In this work we have presented our analysis of the galaxy-galaxy merger rate within the context of the empirical model for galaxy formation \textsc{emerge}. This model connects galaxy growth directly to the halo growth in $N$-body simulations using simple relations constrained by a suite of observables \citep[see][for more details]{moster2018}. We investigated a range of properties associated with galaxy merger rates, including: scaling with stellar mass and mass ratio, the relationship between the merger rates and observed galaxy pairs, the merging history of large galaxies, and the role of merging in galaxy quenching. We also presented a brief comparison of our results to other theoretical models.

We find a galaxy merger rate density $\Gamma$ that increases with redshift until $z\approx1.5$, followed by a sharp decline in the rate towards higher redshift. This general trend holds for the three mass bins we explored between $9\leq \log_{10}(m*)<12$. For the merger rate per galaxy $\mathfrak{R}$, unlike previous works, we do not exhibit universal power law scaling with redshift. We find that merger rates show an excess over a power-law scaling for $z\gtrsim3$. This effect is most apparent when determining merger rates based on descendant galaxy mass.
Generally mergers occur at a higher frequency with increasing galaxy mass. When exploring how merger rates scale with mass ratio, we find that the largest galaxies are biased to experience major mergers and subsequently show an enhanced major merger rate compared to lower mass galaxies. This effect can be seen even for $\mu \lesssim 10$. This effect is a departure of the nearly mass independent scaling in the numerically derived halo-halo merger rate. We conclude that the self similarity shown in halo-halo mergers is broken through the complex connection between galaxies and their haloes. In this view a major halo-halo merger occurring along the high mass slope of the SHMR are likely to result in an eventual major merger in stellar mass as well. Conversely, a major halo-halo merger along the low mass slope of the SHMR is generally more likely to eventually produce a very minor merger in galaxy stellar mass. The influence of the SHMR on major merger rates has been explored in the past \citep[]{Stewart_2009, Hopkins2010}, due to advance in observations and modelling techniques we are able to show that such a phenomenon causes a break from a power-law redshift scaling, particularly at intermediate masses.

We show that our model produces galaxy pair fractions consistent with observations out to high redshift. Despite general agreement in the redshift scaling of the pair fraction, there remains considerable tension between observation and theoretical predictions. Discrepancies in methodologies make a direct comparison between models and even between observations difficult. Subsequently, predictions have not converged to a single functional form for pair fraction evolution. Our model can most reasonably be fit with a power-law exponential form, consistent with the observations from \citet[]{Jiang2014, Man_2016, Mundy2017, Ventou2017, mantha2018}.Our results best match those of \citet[]{Ventou2017}, who employ a redshift proximity criterion most similar to ours owing to their use of spectroscopic redshift information. Following the pair selection criteria of \citet[]{Ventou2017} we find a pair fraction that ranges between $2$ per cent and $7$ per cent.

Differences are further compounded when translating observed (simulated) pair fractions into galaxy-galaxy mergers rates due to the necessity of a well defined observation timescale $T_{\mathrm{obs}}$. When using published values for $T_{\mathrm{obs}}$ \citep[]{lotz2011}, we find a merger rate that over predicts our model intrinsic results by more than a factor of 2. Further, we find that utilizing an observation timescale that scales $\propto(1+z)^{-2}$ \citep[]{Snyder2017} over-predicts our predicted merger rates by nearly an order of magnitude.
Converting our simulated pair fractions to merger rates is most consistent with a linearly evolving observation time scale $T_{\mathrm{obs}}=w(1+z)+b$. The results presented here are a first pass at confining the observation timescales through our model. A more complete analysis taking into account more accurate mock observable implementations, and complete description of the correction factor $C_{\mathrm{merge}}$ is required before more definitive statements can be made. However, we do not expect future work within this model to produce the strong scaling seen by \citet[]{Snyder2017}. Additionally, more complete inspection of the pair fraction sensitivity to observables and model variations is necessary. Other recent works have shown the pair fraction sensitivity to the SHMR \citep[]{Grylls2020}, understanding how changes in these statistical relations impacts observed pair fractions is vital to understanding galaxy clustering and merging timescales at high redshift.

Our model predicts merger rates that are consistent with other theoretical models. However, within the range of previous works, our results tend to sit lower than the average. The merger rates produced from our model are in closest agreement with \citet[]{bower2006}. Additionally, our mass dependent bias towards major mergers is an effect absent in some other models \citep{Rodriguez-Gomez2015}. A more complete discussion of the intricacies and differences in these models can be found in \citet[]{hopkins10b}.

Additionally, we also explored a diverse set of model variations to determine the merger rate sensitivity to a range of model options. We show that our newly implemented model for galaxy stripping does not result in overly aggressive satellite loss, suppressing the merger rate. Further, we set reasonable upper bounds on the expected merger rate in the context of our model by systematically varying the treatment of orphan galaxies. In particular we illustrate that application of dynamical friction delay at subhalo destruction does not result in a prolonged merger timescales compared to application at $R_{\mathrm{vir}}$ for the simulation volume we tested. We also verified that major satellites are not spending spending too much time in the orphan phase by merging galaxies directly at the time of subhalo destruction. we also showed that our model is robust to large increase in particle resolution by running our model on top of the Illustris TNG100-Dark merger trees. Ultimately none of the reasonable model changes we tested resulted in a substantial change to major merger rate normalisation or redshift evolution. We can conclude that our results are reliable to within a factor $\sim2$ for the current model and parameter set.

When testing model variations we also explored galaxy merger rates derived from average halo-halo merger rates. This test reproduced the methods described by \citep{Hopkins2010}, with updates to better reflect our improved fit to the relevant observables. We found that these halo averaged merger rates over predict galaxy merger rates by as much as an order of magnitude compared with rates derived from complete galaxy merger trees.

We also explored the merging history of the $z=0$ galaxy population to determine what role mergers play in the buildup of stellar mass. Our model shows that galaxies with $\log_{10}(m/M_{\odot})\lesssim11$ grow almost entirely through \textit{in-situ} star formation, with accreted material accounting for $\leq10$ per cent of the total stellar mass. Furthermore, this accreted material is overwhelmingly deposited in minor or mini mergers, with $\geq 90$ per cent of accreted stellar mass attributed to mergers with $\mu\geq4$. For more massive galaxies $\log_{10}(m/M_{\odot})\gtrsim 11$, galaxy-galaxy mergers play a critical role in the buildup of mass. In these galaxies, accreted material accounts for as much as $80$ per cent of the total stellar mass for some galaxies. In these cases the stellar mass is largely deposited through major mergers, where as much as $90$ per cent of the total accreted mass is delivered through mergers with $\mu\leq4$.

Our results indicate that \textsc{emerge} can accurately predict the galaxy merger rate out to high redshift. We are able to not only compute the cosmic merger rate of galaxies but explore the individual merging history of each galaxy in our simulation volume. Additionally, we have shown that mock observables derived from our simulation agree with recent observations within a factor $\sim2$, at least when spectroscopic data is available \citep{Ventou2017}. Despite these successes, additional work is needed to narrow the gap between model predictions and observations. In particular, the mass and redshift dependencies of the observation timescales for close galaxy pairs has to be addressed in more detail. Lastly, at the time of the analysis our model did not include any information about the gas properties or orbital configurations of merging systems. These additional details may be necessary to form a complete understanding of the impact of galaxy merging on the star formation properties, and radial distribution of mass in observed galaxies.

\section*{Acknowledgements}

We thank all authors who provide their data in electronic form. JAO thanks Andrew Hearin for his quick implementation of desired functionality into \textsc{halotools} \citep{Hearin2017}, as well as Shy Genel and Austen Gabrielpillai for making the TN100-Dark \citep{Springel2018} \textsc{rockstar} \citep{rockstar} trees available to us. We are also grateful to Rhea-Silvia Rhemus, Ulrich Steinwandel, Felix Schulze, Thierry Contini, and Peter Behroozi for enlightening discussions. The cosmological simulations used in this work were carried out at the Odin Cluster at the Max Planck Computing and Data Facility in Garching. BPM and JAO acknowledges an Emmy Noether grant funded by the Deutsche Forschungsgemeinschaft (DFG, German Research Foundation) -- MO 2979/1-1. TN acknowledges support from the DFG Origins Cluster.

\section*{Data availability}
The source code of \textsc{emerge} and sample galaxy merger trees are available on \url{https://github.com/bmoster/emerge}. The derived data and analysis scripts used in this article can be found at \url{https://github.com/jaoleary}.




\bibliographystyle{mnras}
\bibliography{references} 






\bsp	
\label{lastpage}
\end{document}